\documentclass[showpacs,showkeys,preprintnumbers,amsmath,amssymb,times,graphicx]{revtex4}

\usepackage{graphicx} 
\usepackage{dcolumn}
\usepackage{bm}

\usepackage{color}
\usepackage{amsmath}
\usepackage{amsfonts}
\usepackage{amssymb}
\usepackage{amsbsy}
\usepackage[figuresright]{rotating}
\usepackage[font=small,labelfont=bf]{caption}
\usepackage{natbib}

\def\3{\ss}

\def\q0{\phantom{1}}

\def\thetabn{\Theta_{Bn}}

\def\parp2{\frac{\partial^{2}}{\partial p^{2} }}

\def\CL{\footnotesize CLUSTER}
\def\SL{\footnotesize SLAMS}

\def\m21{$2^{\circ}\times 1^{\circ}$}

\def\ts{\thinspace}

\def\ne8{Ne\ts{$\scriptstyle {\rm VIII}$} }

\def\CL{\footnotesize CLUSTER}
\def\SL{\footnotesize SLAMS}

\newcommand{\asr}{{Adv. \ Space. \ Res.}}

\newcommand{\ag}{{Ann.\ Geo\-phy\-si\-c\ae}}

\newcommand{\grl}{{Geo\-phys.\ Res.\ Lett.}}

\newcommand{\jgr}{{J.\ Geo\-phys.\ Res.}}

\newcommand{\pss}{{Planet.\ Space Sci.}}

\newcommand{\ssr}{{Space Sci.\,Rev.}}

\newcommand{\pf}{{Phys. Fluids}}

\newcommand{\pop}{{Phys.\ Plasmas}}

\def\ion[#1 #2]{#1\,{\sc #2}}
\def\lamb[#1]{#1\,{\AA}}
\def\serts89{SERTS-89}
\def\fe12{Fe\,{\sc xii}}

\def\mb[#1]{\makebox[0.15cm][l]{#1}}

\begin{document}

\title{Fundamentals of  Non-relativistic Collisionless Shock Physics: \\  V. Acceleration of Charged Particles}

\author{R. A. Treumann$^\dag$ and C. H. Jaroschek$^{*}$}\email{treumann@issibern.ch}
\affiliation{$^\dag$ Department of Geophysics and Environmental Sciences, Munich University, D-80333 Munich, Germany  \\ 
Department of Physics and Astronomy, Dartmouth College, Hanover, 03755 NH, USA \\ 
$^{*}$Department Earth \& Planetary Science, University of Tokyo, Tokyo, Japan
}%

\begin{abstract} A comprehensive review is given of the various processes proposed for accelerating particles by shocks to high energies. These energies are limited by several bounds: the non-relativistic nature of the heliospheric collisionless shocks to which this review restricts, the finite size of these shocks, the finite width of the downstream region, and to the nature of turbulence. In general, collisionless shocks in the heliosphere cannot accelerate particles to very high energies. As a fundamental problem of the acceleration mechanism the injection of see particles is identified. Some mecchanisms for production of seed particles are invoked. Acceleration of electrons begins to uncover its nature. The following problems are covered in this chapter: 1. Introduction -- first and second order Fermi acceleration,  2. Accelerating ions when they are already fast, diffusive acceleration, convection diffusion equation, Lee's self-consistent quasilinear shock acceleration model, 3. Observations, 4. The injection problem, ion surfing, test particle simulations, self-consistent shock acceleration simulations, downstream leakage, trapped particle acceleration,  5. Accelerating electrons, Sonnerup-Wu mechanism, Hoshino's electron shock surfing on quasi-perpendicular shocks, quasiparallel shock surfing.
\end{abstract}
\maketitle 

\section{Introduction}\noindent
Interaction of fast particles with shocks has been proposed by {\it Enrico} \cite{Fermi1949} as a simple natural mechanism of pushing {\it charged}-particle energies up to very high values, such high that the respective particles in the Universe can appear as Cosmic Rays. 

Fermi's mechanism is based on the insight that the particle energy while being a constant of the particle motion is not an invariant with respect to coordinate transformations. For instance, when sitting on a fast (classical) particle one does not feel its velocity. What is felt is just its {\it Zitterbewegung}, i.e. its jitter motion caused by its internal oscillations, and the acceleration and deceleration it experiences when interacting with fields that are external to it, large scale external electric or magnetic fields, and the proper fields of the surrounding particles. The latter interaction is what causes its thermal velocity $v_{e,i}$, or in terms of energy, its temperature $T_{e,i}$, taken in energy units -- J or eV --, where the indices $e,i$ refer to either electrons or ions. \index{Fermi, Enrico} \index{acceleration!Fermi}

However, seen from the rest frame, one adds to these energies the translational energy of the particle. Thus, a particle that passes from downstream across the shock to upstream, is reflected there, and returns to downstream, gains the  velocity difference between the upstream and downstream media and thus has gained an energy that (non-relativistically) is of the order of $\Delta{\cal E}=2m_{e,i}V|v_\||\sim2m_{e,i}V_{A1}{\cal M}|v_\||$, and $\Delta |v_\||=2V$, where $V$ is the velocity of the scatterer (`mirror') \citep{Jokipii1966}. The second part of this expression holds only in an approximate sense as we have not included the angular dependence of the energy conserving scattering process and the obliqueness of the particle velocity. A more precise argument \citep[originally given by][]{Bell1978} can be found in \cite{Drury1983}. A downstream particle of initial relativistic velocity $\beta\cos\theta_{vn}>V_2/c$ and pitch angle $\theta$ in the downstream region will cross the shock upstream into the region moving with speed $V_1$, become scattered in the flow (thereby conserving energy), and returns downstream with pitch angle $\theta^\prime_{vn}$. In the downstream frame its total (relativistic) energy gain $\Delta{\cal E}={\cal E}^\prime-{\cal E}$, with ${\cal E}=p c/\beta$ the relativistic energy, is 
\begin{equation}
\frac{\Delta{\cal E}}{{\cal E}}=\frac{\Delta V}{c}\frac{\beta\cos\theta_{vn}-\beta^\prime\cos\theta_{vn}^\prime}{1+\beta^\prime\Delta V\cos\theta^\prime_{vn}/c}, \qquad \Delta V= V_1-V_2
\end{equation}
which is positive because for returning particles $\pi/2<\theta_{vn}$ and $0<\theta_{vn}'<\pi/2$. Here $\Delta V$ is the velocity difference between the upstream and downstream scatterers (note that these, strictly speaking, usually move at velocities that are different from the respective flows!). Thus the particle energy increases monotonically for multiple shock crossings, as had been suggested by \cite{Fermi1949}.

\begin{figure}[t!]
\hspace{0.0cm}\centerline{\includegraphics[width=0.70\textwidth,clip=]{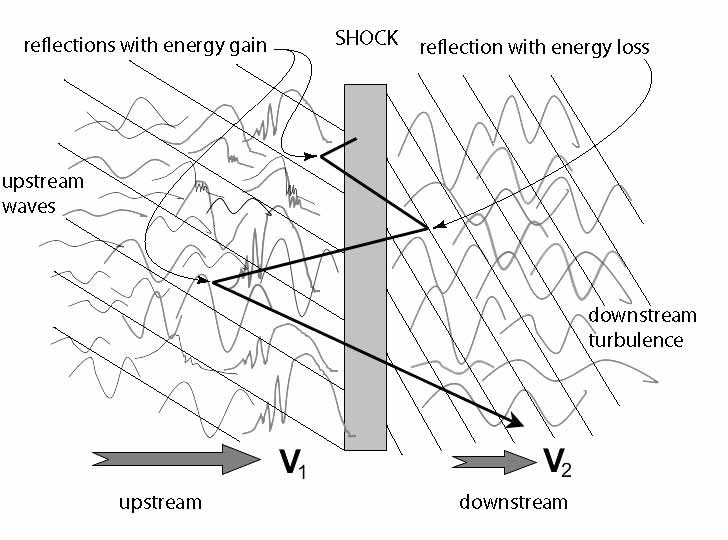}}
\caption[Reflection]
{\footnotesize The schematic representation of the acceleration mechanism of a charged particle in reflection at a quasi-parallel supercritical shock. on the left of the shock is the upstream plasma flow of velocity ${\bf V}_1 \gg{\bf V}_2$ much larger than the downstream velocity. It contains the various upstream plasma modes: upstream waves, shocklets, whistler, and pulsations. The average magnetic field is inclined at  small $\thetabn<45^\circ$ against the shock normal. On the downstream side the plasma is flowing slowly and contains the downstream turbulence. The energetic particle that is injected at the shock to upstream is reflected in an energy gaining collision with the upstream waves, moves downstream where it is reflected in an energy loosing collision back upstream. It looses energy because it overtakes the slow waves, but the energy loss is small. Returning to upstream it is scattered a second time again gaining energy. In this way its initially already high energy is successively increased until it escapes from the shock and ends up in free space as an energetic Cosmic Ray particle. The energy gain is on the expense of the upstream flow which is gradually retarded in this interaction. However, the number of energetic particles is small and the energy gain per collision is also small. So the retardation of the upstream flow is much less than the retardation it experiences in the interaction with the shock-reflected low energy particles and the excitation of the upstream turbulence.  }\label{chap6-fig-refl}
\end{figure}

Clearly, in one single reflection cycle back upstream and back downstream the particle does not pick up a large amount of energy and momentum. Thus, in order for the acceleration to be efficient, a large number of shock crossings and reflections back and forth is required. This is schematically shown in Figure\,\ref{chap6-fig-refl}. Shock particle acceleration therefore depends on the scattering process which is clearly a stochastic process, depending on the presence of scattering centres upstream and downstream and on the probabilistic nature of the changes in the scattering angle. It also assumes that the scattering conserves energy, i.e.  the scattering is assumed elastic such that the scattered particles are practically not involved into any form of dissipation of their energy of motion. In particular they should not become involved into excitation of instabilities which consume part of their motional energy. The only actual dissipation that is allowed in this process is dissipation of bulk motional energy from where the few accelerated particles extract their energy gain. On the other hand `dissipation' is also attributed to direct particle loss by either convective transport or the limited size of the acceleration region. This mechanism therefore works until the gyroradius of the accelerated particle becomes so large that it exceeds the size of the system or, otherwise,  in a very extended plasma -- for instance in the environment of an interstellar travelling shock --, until the energy of the particle becomes so large that the back-scattering of the particle becomes ineffective. 
\index{Cosmic Rays!spectrum of}

The dependence on the gyroradius imposes a severe limitation on the acceleration mechanism. In order to experience a first scattering, i.e. in order to being admitted to the acceleration process, the particle must initially already posses a gyroradius much larger than the entire width of the shock transition region. Only when this condition is given, the shock will behave like an infinitesimally thin discontinuity separating two regions of vastly different velocities such that the particle when crossing back and forth over the shock can become aware of the bulk difference in speed and take an energetic advantage of it. This restriction is indeed very rigid as it rules out any particles in the core of the upstream inflow distribution from participation in the acceleration process. In fact, in order to enter the Fermi shock-acceleration mechanism a particle must be pre-accelerated or pre-heated until its gyroradius becomes sufficiently large. This condition poses a problem that has not yet been resolved and to which we will return several times in the present chapter.  \index{Davis, Leverett}

Being aware of this difficulty, {\it Leverett} \cite{Davis1956} \index{Davis, Leverett} extended and modified Fermi's mechanism to include a medium with many diffusely distributed scattering centers in relative motion. It is clear that in this process the resulting accelerated particle distribution will have been scattered into all angles; it will be about isotropic or ``diffuse". Thus this mechanism is now known as the ``diffusive"  or ``first order" Fermi-acceleration mechanism (leaving the original shock-acceleration ``second-order" Fermi acceleration), which we will discuss in application to the shock environment. However, this mechanism suffers from the same disease as Fermi's second-order mechanism as acceleration of thermal particles is inefficient as well. Reviews of the shock Fermi acceleration mechanisms working under different space and astrophysical conditions are found among many others in the papers and books  of \cite{Jokipii1971}, \cite{Drury1983}, \cite{Forman1985},  \cite{Blandford1987}, \cite{Jones1991}, and \cite{Schlickeiser2002}. 
\begin{figure}[t!]
\hspace{0.0cm}\centerline{\includegraphics[width=0.70\textwidth,clip=]{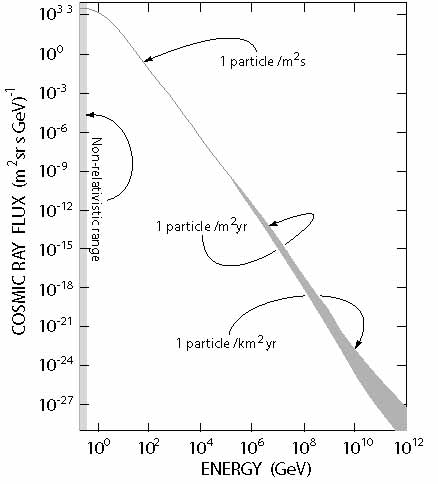}}
\caption[FS waves disprel]
{\footnotesize The Cosmic Ray spectrum. The nonrelativistic range is indicated in shading at the far left of the figure \citep[after][]{Cronin1997}. The spectrum is approximately power law with breaks in the power at several energies. At the highest energies the count rate is very low and the shape of the spectrum is therefore quite uncertain. However, the impression is that the spectrum follows an approximate power law over ten decades. }\label{chap6-fig-CR}
\end{figure}\index{Cosmic Rays!spectrum of}

The vivid interest of astrophysics into shock acceleration has in the past six decades led to an enormous activity in this field and the publication of an uncountable number of papers, which liberates us from the obligation of producing an exhaustive review of the various approaches and achievements in shock acceleration theory and application. In addition, Cosmic Rays are of much higher energy than can be generated in the heliosphere referring to the non-relativistic shocks it contains. Figure\,\ref{chap6-fig-CR} shows the Cosmic Ray spectrum as is known today from various observations. The belief in Cosmic Ray acceleration by shocks is large fuelled by the spatial isotropy of Cosmic Rays as well from its approximate power law shape over wide ranges of the spectrum even though the spectrum exhibits several breaks in this shape (see the figure) and becomes quite uncertain at extremely high energies. However, Cosmic Rays require highly relativistic or even ultrarelativistic shocks \citep[cf, e.g,][]{Waxman2006}. Thus the contribution of heliospheric shock acceleration is quite naturally restricted to the range of weakly relativistic particles and to the investigation of particle acceleration by measuring energetic particle spectra {\it in situ} the shock environment. These measurements can then be compared with theory and in the first place numerical simulations in order to select the relevant acceleration models for medium energy particles ($<$ GeV ions and $<$ MeV electrons). 

In addition, because of the availability -- or at least the occasional availability -- of collisionless shocks in space, like planetary bow shocks, travelling interplanetary shocks, corotating interaction regions, coronal shocks and the heliospheric terminal shock, one of the most interesting questions in shock acceleration theory can be treated. This is the above mentioned complex of questions that are related to the so-called shock particle\index{process!shock injection} {\it injection problem}: Which of the various mechanisms is capable of accelerating ions and electrons out of the main streaming thermal plasma distributions to energies high enough that they can become injected into the cycle of the shock-Fermi acceleration machine? Theory has so far been unable to ultimately answer this question. However, a number of sub-processes acting in the shock have in the past been proposed of which it is believed that some of them are indeed capable of contributing to answering this question. This problem does not directly stimulate astrophysical interest as it is believed that in the huge astrophysical objects with the available high energies sufficiently many particles will always have sufficiently high energy for initiating the Fermi process. Here another problem awakens attention even when the shocks are non-relativistic: this is the question what happens to a shock, if it is exposed to a substantial density of energetic particles, particles that have undergone Fermi acceleration and fill all the space upstream and downstream of the shock. These particles are believed to modulate the shock, transforming it into a energetic particle (or Cosmic Ray) {\it mediated} shock wave.\index{process!shock mediation}\index{shocks!Cosmic Ray mediation} We are not going to treat this problem here as in the heliosphere there is presumably only one single shock that may be subject to weak modulation by the Anomalous Cosmic Ray  component that is present in the heliosphere, the Heliospheric Terminal Shock,\index{shocks! Heliospheric Terminal} which we will briefly treat in passing in the second part of this volume. 
\begin{figure}[t!]
\hspace{0.0cm}\centerline{\includegraphics[width=0.95\textwidth,clip=]{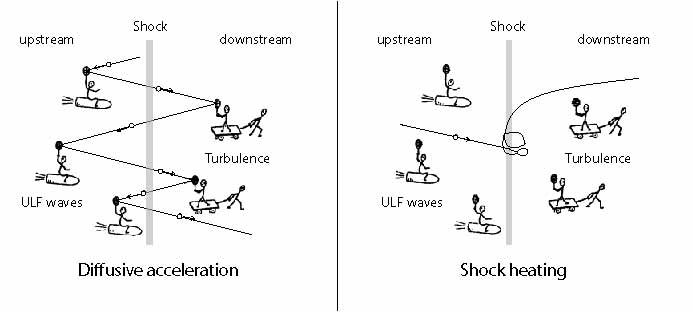}}
\caption[FS waves disprel]
{\footnotesize Cartoon of the diffusive shock acceleration (left) and shock heating mechanisms \citep[after][after an original sketch by M. Scholer]{Hoshino2001}. In diffusive shock acceleration the particle is scattered around the shock being much faster than the shock. The requirement is the presence of upstream waves and downstream turbulence or waves. In shock heating the particle is a member of the main particle distribution, is trapped for a while at the shock and thereby thermalised and accelerated until leaving the shock. }\label{chap6-fig-cartoon}
\end{figure}\index{Cosmic Rays!spectrum of}

\section{Accelerating Ions When They Are Already Fast}
\noindent When dealing with the acceleration of particles by shocks, the physics of the shock stands back and is not of large interest. The shock appears as a boundary between two independent regions of different bulk flow parameters which are filled with scattering centres for the particles as sketched in Figure\,\ref{chap6-fig-refl} (see also the cartoon in Figure\,\ref{chap6-fig-cartoon}). These scattering centres are assumed to move at speeds not very different from the speeds of the bulk flows to both sides while being capable of scattering even high energy particles in angle. The scattering process is in most cases modelled by some dependence of the mean free path of the particles on momentum, or it is modelled according to some self-consistent quasilinear velocity space diffusion coefficient that takes account of the particle distribution function. Either method is not completely self-consistent since the formation of the shock is not taken into account and since the injection process remains unsolved. We will therefore only briefly go through the theory and later refer to the simulational approach to the shock acceleration problem. 
\subsection{Diffusive acceleration}\index{acceleration!diffusive}
\noindent The multiple transitions of a fast particle back and forth across the shock under the assumption that the scattering is a stochastic process suggests a diffusive approach to the acceleration mechanism. Before briefly going into detail of the diffusive treatment, we follow \cite{Bell1978} in giving a simple heuristic argument for the resulting particle distribution being power law in momentum or velocity.

\subsubsection{The heuristic argument} 
\noindent Assume that the particle that is to be scattered has already sufficiently high initial velocity $\Delta V\ll v_{\rm in}$ or initial moment $p_{\rm in}$. In this case the relative change of the particle moment $p_j$ in one scattering $j$ is $\Delta p_j/p_j=(\cos \theta_j-\cos\theta_j')\Delta V/v_j$, which by ensemble averaging over the scattering angle in the interval $0\leq \theta\leq\frac{\pi}{2}$ simply produces a factor $\frac{2}{\pi}$ per each cosine, yielding $\langle\Delta p_j/p_j\rangle=\frac{4}{\pi}\Delta V/v_j$ for the average change in momentum during the scattering process. (Note that averaging over the angular distribution implies complete isotropy, which is a crucial assumption in diffusive acceleration theory leaving us just with the momentum dependence!) Now assume that the particle of initial moment $p_{\rm in}$ has indeed experienced a large number $J$ of scatterings before being injected away from the shock into space. Then the final particle moment becomes 
\begin{equation}
\frac{p_J}{p_{\rm in}}=\prod\limits_{j=1}^{j=J}\left(1+\frac{4}{\pi}\frac{\Delta V}{v_j}\right)\simeq \exp\left(\frac{4}{\pi}\sum_{j=1}^{j=J}\frac{\Delta V}{v_j}\right)
\end{equation}
where the expression in brackets is understood as the expansion of the exponential for small argument. When the loss of particles from the downstream  region 2 is mainly convective, then the probability of loss is the ratio of downstream flux to entering flux from upstream, $NV_2 /\frac{1}{4}Nv$, and the probability for a scattering to happen before the particle is convectively transported away is $P_j=1-NV_2 /\frac{1}{4}Nv_j$. Thus the $J$s probability, i.e. the  total probability at the last scattering in the Fermi cycle is 
\begin{equation}
P_{\geq J}=\prod_{j=1}^{j=J}\left(1-\frac{4V_2}{v_j}\right)\simeq\exp\left(-4\sum_{j=1}^{j=J}\frac{V_2}{v_j}\right)
\end{equation}
Writing this in terms of the momentum by eliminating the sum in the exponential with the help of the above expression for $p$ (dropping the large arbitrary index $J$), the probability of a particle to reach its final momentum is found to be
\begin{equation}
P_{\geq p} =\exp \left\{ -\frac{\pi V_2}{\Delta V} \ln\frac{p}{p_{\rm in}}\right\}= \left(\frac{p}{p_{\rm in}}\right)^{-\frac{\pi V_2}{\Delta V}} \propto N_{\geq p}
\end{equation}
This expression is a power law for the probability, and the power is a constant which is given by the ratio of the convective loss probability to the change in $\ln p$ per scatter and depends just on the shock compression ratio. This result is also the number density $N_{\geq p}$ of particles with momenta larger $p$. Since this can be expressed as the integral $N_{\geq p}=4\pi\int_{\geq p} p^2F(p){\rm d}p$ over the isotropic distribution function $F(p)$, one obtains for the latter
\begin{equation}
F(p)=-\frac{1}{4\pi p^2}\frac{{\rm d}N_{\geq p}}{{\rm d}p}\simeq \frac{\pi V_2}{\Delta V}\frac{N_{\geq p}}{4\pi p_{\rm in}^3}\left(\frac{p}{p_{\rm in}}\right)^{-\frac{\pi V_1}{\Delta V}\left[1-\left(1-\frac{3}{\pi}\right)\frac{V_2}{V_1}\right]}
\end{equation}
The momentum distribution is a power law $F(p)\simeq (p/p_{\rm in})^{-\alpha}$ with $\alpha\approx \pi V_1/\Delta V$ \citep{Bell1978,Drury1983,Forman1985}. (Fermi's second-order acceleration yields essentially the same formula, however with $\alpha=1+\tau_{\rm acc}/\tau_{\rm esc}$ being the ratio of acceleration time to escape time. The relative energy gain in one collision is then $\sim |V/c|$ for the scatterer moving with speed $V$, while energy-increasing collisions are favoured and introduce another factor $V/c$. Thus the acceleration time is second order $\tau_{\rm acc}\sim c^2/V^2$ -- from where the term ``second-order" Fermi originates.)

The above power law solution of the distribution function has two deficiencies: it is valid only for large momenta $p>p_{\rm in}$ and, thus, does not describe the low energies nor the way how particles can cross the $p_{\rm in}$-boundary in momentum space in order to become injected; in addition, however, it does not describe particle losses through a distant boundary. The former problem is of principal nature and cannot be solved by a simple acceleration theory; its solution must take care of the entire shock formation process and particle dynamics, which in a simple acceleration like the one described here theory is not included. The loss problem can be accounted for by simply assuming a probability of loss. This can be done in two ways, either assuming that the particles, when crossing a distant boundary located at $x=\Delta_{\,\rm d}$ are considered lost and do not return. This probability is simply $P_{\,\,\rm L}=r_{ci}/ \Delta_{\,\rm d}$, i.e. the ratio of gyroradius to  width. Or, one also accounts for an additional convective return to the shock. Then it becomes the product of the loss and convective return probabilities $P_{\,\,\rm L}=(r_{ci}/\Delta_{\,\rm d})(r_{ci}v/V_1\Delta_{\,\rm d})$, which is a reduction of the escape probability \citep{Jones1991}. The effect of the loss is an exponential cut-off of the distribution which is most severe if the convective return is neglected:
\begin{equation}
F(p)\to F(p)\exp \left(-\frac{3 p^2/2m_i}{ZeB\,\Delta V \Delta_{\,\rm d} }\right)
\end{equation}
which shows that the cut-off depends on the energy to charge ratio $p^2/2m_iZe$.  

\subsubsection{Fokker-Planck diffusion equation}\index{equation!Fokker-Planck} 
\noindent Whatsoever the nature of the scatter is, the assumption of a stochastic process implies that the basic equation that governs the process is a phase space diffusion equation of the kind of a Fokker-Planck equation
\begin{equation}
\hspace{-0.3cm}\frac{\partial F({\bf p, x},t)}{\partial t}+{\bf v}\cdot\nabla F ({\bf p, x},t) =\frac{\partial}{\partial{\bf p}}\cdot \textsf{D}_{{\bf pp}}\cdot\frac{\partial F({\bf p, x},t)}{\partial{\bf p}} , \qquad \textsf{D}_{{\bf pp}} =\frac{1}{2}\left\langle\frac{\Delta{\bf p}\Delta{\bf p}}{\Delta t}\right\rangle
\end{equation}
where $\Delta{\bf p}$ is the variation of the particle momentum in the scattering process which happens in the time interval $\Delta t$, and the angular brackets indicate ensemble averaging. $\textsf{D}_{{\bf pp}}$ is the momentum space diffusion tensor. It is customary to define $\mu=\cos\alpha$ as the cosine of the particle pitch angle $\alpha$ and to understand among $F(p,\mu)$ the gyro-phase averaged distribution function, which depends only on $p=|{\bf p}|$ and $\mu$. 

The scattering centres are of course waves of phase velocities $v_{\rm ph}\equiv \omega/k\ll v_\|=v\mu$. In this case the diffusion tensor has only the three components $D_{\mu\mu}, D_{\mu p}=D_{p\mu}, D_{pp}$. These can all be expressed through the pitch-angle diffusion coefficient $D_{\mu\mu}$ and are given by the expressions
\begin{equation}
\hspace{-0.3cm}D_{\mu\mu}=\frac{\pi\omega_{ci}}{2}\left.\frac{|k|W_k}{B^2}\right|_{\rm res}(1-\mu^2), \qquad \frac{D_{p\mu}}{D_{\mu\mu}}=\left(\frac{pv_{\rm ph}}{v}\right),\qquad \frac{D_{pp}}{D_{\mu\mu}}=\left(\frac{pv_{\rm ph}}{v}\right)^{\!\!2}
\end{equation}
Here the spectral power density  $W_k$ of the waves that are involved into the scattering has to be taken at the resonant wave number $k_{\rm res}=\mu v/\omega_{ci}$, and the ion cyclotron frequency is given by its relativistic expression $\omega_{ci}=ZeB/\gamma$. 

One immediately realises that pitch angle scattering is the main process at small phase velocities as long as $p$ becomes not too large, because the energy diffusion $D_{pp}$ in scattering is quadratically small in the ratio $pv_{\rm ph}/{v}$ becoming more important only at large momenta $p$. One may use the pitchangle diffusion coefficient to define a pitch-angle scattering frequency as $\nu_\mu=2D_{\mu\mu}/(1-\mu^2)\ll\omega_{ci}$ which, of course, is much less than the cyclotron frequency. The mechanism of acceleration is, hence, a (fast) multi-pitchangle scattering process followed by slow energy diffusion, just as we have suspected from simple logic that multiple crossings of the shock (or otherwise multiple reflection from moving waves) would be required in order to push the particles to high energy.

\subsubsection{Wave spectra involved} 
\noindent The dependence on the wave spectral power density $W_k$ is important. It implies that the entire diffusive acceleration process depends crucially on the assumption on which waves are involved. By the nature of the diffusive approximation of the Fokker-Planck equation these waves must be distributed isotropically in space, a condition that is barely satisfied in the vicinity of curved shock waves like planetary bow shocks in a high Mach number stream like the solar (or stellar) wind. 
the waves are probably in the fast magnetosonic wave band. 
\begin{figure}[t!]
\hspace{0.0cm}\centerline{\includegraphics[width=0.95\textwidth,clip=]{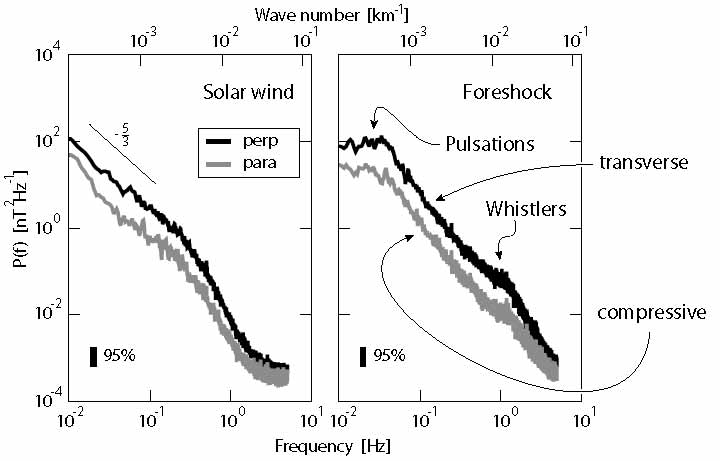}}
\caption[FS waves disprel]
{\footnotesize Power spectral densities for the magnetic fluctuations measured by CLUSTER in the solar wind and foreshock versus time $t$ and wave number $k$.  The spectra distinguish between perpendicular (transverse) and parallel (compressive) fluctuations indicating the strong anisotropy of the fluctuations both in the solar wind (left) and the foreshock (right). The perpendicular fluctuations are more  intense than the parallel fluctuations. In the solar wind the spectra decay approximately like Kolmogorov for low frequencies. In the foreshock the spectra are flat at low frequencies according to the presence of shocklets and pulsations ({\SL}), exhibit a maximum at roughly $\sim0.05$\,Hz which is the frequency range of pulsations corresponding to transverse wavelengths of $\lesssim 10^3$\,km. At higher frequencies and wave numbers the spectra are again similar to Kolmogorov though the range of power law is short and the uncertainty might hide a steeper decay \citep[after][]{Narita2008}.}\label{chap6-fig-narita1}
\end{figure}

A digression in our previous investigation of supercritical shocks reveals that upstream of quasi-perpendicular shocks susceptible wave power is found only in the shock foot region. The transverse size of shock feet is of the order of one (or at most a few) thermal upstream ion gyroradii. Energetic ions entering the upstream region from downstream having been scattered there into the upstream direction presumably completely ignore the foot since it belongs to the shock transition which in this kind of theory is assumed infinitesimally narrow. Both, observations and simulations have shown that farther away from the quasi-perpendicular shock the upstream flow only contains the weak upstream turbulence and possibly a few whistler precursors which have managed to escape some short distance into the upstream flow before being damped out. If the upstream turbulence is incapable of providing a high enough turbulence level, quasi-perpendicular shocks will not become effective diffusive accelerators. In the case of he solar wind, the upstream solar wind turbulence is about Kolmogorov with spectrum $W_k\propto k^{-\frac{5}{3}}$ but is weak \citep{Tu1995,Goldstein1995,Biskamp2003}, in particular when the solar wind is fast, coming from the coronal holes. Higher turbulence levels are observed in slow solar wind streams when it contains a component of travelling discontinuities.\index{turbulence!solar wind} 

On the other hand, upstream of the supercritical quasi-parallel shock there is a substantial level of wave activity, low frequency upstream waves, Alfv\'en-ion cyclotron, whistler waves, shocklets and pulsations ({\SL}). These waves produce a comparably intense though anisotropic turbulence level \citep{Narita2006,Narita2008} which, in addition, decays with increasing distance from the shock. The assumption of an isotropic and homogeneous wave distribution upstream of a quasi-parallel shock is therefore barely realistic. Figure\,\ref{chap6-fig-narita1} shows spectra measured by  {\CL} in the solar wind and the foreshock of the Earth's bow shock wave. In both cases the anisotropy is quite pronounced with the solar wind behaving like  Kolmogorov  at low frequencies/large wavelengths. The foreshock spectrum shows the large contribution of pulsations as well as the contribution of whistlers. The range of power law is shorter than in the solar wind and is possibly also steeper. \index{turbulence!foreshock}

The turbulence in the foreshock is sufficiently intense (see also Figure\,\ref{chap5-fig-narita-turb}) for scattering ions into a nearly diffuse component. However, the theoretical assumption of a single wave component (for instance of purely Alfv\'enic turbulence) is certainly unjustified. Neither shocklets nor pulsations are of this kind of waves; they are localised large amplitude irregular structures the contribution of which to the turbulence is largely unknown. Moreover, their spatial evolution is not accounted for in any of the acceleration theories. On the other hand, because of their large amplitudes and their effect on the magnetic field magnitude and direction they must be rather efficient in the scattering of particles. \index{turbulence!downstream}

The expectation is thus that the scattering is strong at an upstream distance close to the shock and levels out into the ordinary upstream flow scattering efficiency.  But the most severe restriction is the narrowness of the downstream region which downstream of the Earth's bow shock is only one to two Earth radii thick. Once the particle reaches a gyroradius larger than the width of the downstream region the probability for being scattered back upstream must drastically decrease. This spatial restriction must naturally impose severe limits on the backscattering of ions from downstream to upstream thereby cutting off the Fermi cycle. The gyroradius is $r_{ci}=p/ZeB$. For a proton gyroradius of 2\,R$_{\rm E}$ in a field of $30$\,nT the momentum is $p\simeq10^{-19}$ Js/m yielding an energy $\epsilon_i/m_ic^2\simeq1.002$, which corresponds to a maximum energy of $\epsilon_{i{\rm max}}\sim(1-2)$\,MeV. On the other hand the high level of wave activity downstream of the shock compensates partially for its narrowness. Spectral wave energy densities downstream of the quasi-parallel shock are at least two orders of magnitude above those in the foreshock \citep[cf, e.g, the recent {\CL} observations by][]{Narita2008}. Since these enter the diffusion coefficients linearly the scattering probability behind the shock is much higher than from the foreshock.

Hence the question arises, what the shock is good for in acceleration if the downstream region limits the acceleration by being too narrow and the upstream region either containing weak turbulence as in the case of the quasi-perpendicular shock or containing inhomogeneous and anisotropic turbulence like in the case of quasi-parallel shocks. The answer is that shocks should serve as pre-accelerators for energetic particles. Probably they are the sources of the energetic seed particle component for further acceleration. For this they must trap ions and electrons for long enough time in the shock transition. Before discussing the related injection problem we will return to the diffusive shock acceleration, however, in order to briefly complete the picture. 

\subsection{Convective-diffusion equation}\index{equation!convective-diffusion}
\noindent Since convective losses enter the above quasi-liner diffusion equation, the equation can be rewritten into a diffusion equation in real space by taking into account that convection produces a streaming density term ${\bf s}(p)=\int{\bf v}F(p,\mu)(1-\mu^2){\rm d}\mu$ that is added to the diffusion equation. For Alfv\'en waves propagating in both directions along the magnetic field with Alfv\'en speed $V_A$ this term can be written as ${\bf s}= [\textsf{K}\cdot\nabla -\frac{1}{3}{\bf w}p\nabla_p]F(p,\mu)$, where ${\bf w}={\bf V}\pm V_A{\bf B}/B$ is the velocity relative to the Alfv\'en waves which are the presumable scatterers. The new quantity $\textsf{K}$ that appears in this contribution is the spatial diffusion tensor that describes the spatial diffusion of the energetic particle as consequence of the combined pitch angle scattering and convection. In the simple model where the waves propagate just parallel and antiparallel to the magnetic field this diffusion tensor is given by
\begin{equation}\label{chap6-eq-difftens}
\textsf{K}=\left(
\begin{array}{lcr}
\kappa_\| & 0  & 0  \\
0  & \kappa_{\perp,1}  & -\kappa_A  \\
0  &  \kappa_A & \kappa_{\perp,2}  
\end{array}
\right), \qquad \kappa_\|\approx \frac{1}{3}\lambda v, \qquad |\kappa_A|\approx \frac{1}{3}vr_{ci}
\end{equation}
These expressions hold under the assumtpion that $\nu_\mu/\omega_{ci}\ll 1$, and then $|\kappa_A|/\kappa_\|\ll 1$ as well. Usually the perpendicular diffusion $\kappa_\perp$ is also neglected. 

It is clear that this kind of theory is only approximative. The assumption that the scattering waves are Alfv\'en waves is very problematic when referring to our present knowledge on the kind of waves that are present in the vicinity of the shock both upstream and downstream. It is also problematic in view of the solar wind turbulence which is far from being purely Alfv\'enic. In addition, the assumption of purely parallel propagation is barely correct. Pulsations do not propagate parallel to the magnetic field. The waves in the quasi-perpendicular shock foot, which are versions of the electromagnetic modified two stream instability, propagate essentially perpendicular to the field. \cite{Jones1990} has extended the theory for including oblique propagation while still restricting to Alfv\'enic type waves. The analytical expressions obtained are very involved in this case inhibiting an analytical treatment anyway. Finally the neglect of the transverse diffusion of energetic particles is questionable as well. Transverse diffusion is first order in $pv_{\rm ph}/v$ as we have seen and can hardly be neglected, therefore. However, since all these considerations affect only the acceleration of relatively low energy particles, which are affected by the presence of the shock, the deficiencies of the theory are relatively unimportant for high energy particles in which astrophysics is interested. But when dealing with shock mediation and the injection problem they should become crucial.

In the above approximation and with the given expressions for the streaming density function ${\bf s}$ and the spatial diffusion tensor $\textsf{K}$ the isotropic diffusive particle acceleration equation becomes
\begin{equation}
\hspace{-0.3cm}\frac{\partial F}{\partial t}+\nabla\cdot{\bf s}+\frac{1}{p^2}\nabla_p\left[p^2\textsf{J}(p)\right]= {\rm S-L}, \qquad \textsf{J}(p)=\left[{\frac{1}{3}}{\bf V}\cdot\nabla-D_{pp}\nabla_p+{\dot p}\right]F(p)
\end{equation}
This equation is still a quasi-linear equation with the quasilinear phase space streaming current density function $\textsf{J}(p)$. It includes the convective streaming which comes in simply because of the difference in the velocities to both sides of the shock which inhibits transformation to a comoving coordinate system. Such an equation had first been given by \cite{Parker1965} and \cite{Jokipii1971}. The functions on its right include possible sources S and losses L of energetic particles other than convection. 

For more elaborate approximative convective diffusion equations and their numerical or analytical solutions with inclusion of various other wave modes one may consult the extended literature. However, the problem lies not in the mathematical solution of those more complicated equations. It is buried in the fact that the acceleration of Cosmic Rays to high energies can be understood only on the basis of relativistic or even ultra-relativistic shock theory. Non-relativistic shock theory on the other hand should serve to illuminate the initial injection of energetic particles into the acceleration process. Whether this can be treated with the help of approximative diffusion equation is a question that might be doubted. One probably needs to refer to numerical simulation work in two or three spatial dimensions of self-consistent shock generation including the acceleration of particles.

\subsubsection*{One-dimensional case}
\noindent The above convection-diffusion equation including transverse diffusion \citep[following][]{Jones1990} taken in one-dimensional real space and under stationarity (i.e. for asymptotic times $t\to\infty$) becomes  
\begin{equation}
\frac{\partial}{\partial x}\left[V(x)F(x,p)-\kappa\frac{\partial F(x,p)}{\partial x}\right]=\frac{1}{3}\left[\frac{\partial V(x)}{\partial x}\right]\frac{\partial}{\partial p}[pF(x,p)]
\end{equation}
where $\kappa=\kappa_\| \cos^2\thetabn +\kappa_\perp \sin^2\thetabn$ is the projection of the diffusion coefficient onto the shock normal direction. When integrating this equation with respect to the spatial coordinate from upstream (index $u$) infinity $x\to-\infty$ to downstream (index $d$) infinity $x\to\infty$ one may apply the boundary conditions $F(x\to-\infty,p)=F_u(p), F(x\geq 0,p)=F_d(p)$, and $\partial_xV(x)=-\Delta V\delta(x)$ where as usual $\Delta V=V_u-V_d$. Then
\begin{equation}
V_dF_d(p)-V_uF_u(p)=-\frac{1}{3}\Delta V\nabla_p\left[pF_d(p)\large\right]
\end{equation}
It is convenient to define the shock compression ratio $\rho_{sh}= V_u/V_d>1$ with the help of which this equation becomes an inhomogeneous ordinary differential equation for the distribution $F_d(p)$ far downstream $p(\partial F_d/\partial p) +\alpha F_d=(\alpha +2)F_u$, where $\alpha=(\rho_{sh}+2)/(\rho_{sh}-1)$. The solution of this equation is simply
\begin{equation}
F_d(p)=(\alpha+2)p^{-\alpha}\int_{p_{\rm in}}^p\,p'^{(\alpha-1)}F_u(p') {\rm d}p' + Cp^{-\alpha}
\end{equation}
$C$ is an arbitrary integration constant of the homogeneous equation which is put to zero in order to prevent an infrared catastrophe at $p\to 0$. The lower bound $p_{\rm in}$ on the integral is required already large in order to satisfy  the condition $v\gg V$ of validity of the entire diffusive acceleration theory. $p_{\rm in}$ is the momentum of the particles which are injected into the acceleration cycle. Assuming that the injection function $F_u(p)\sim F_u(\epsilon)\sim\delta(\epsilon-\epsilon_{\rm in})$ is a $\delta$-function, one obtains from here the differential energy flux of particles as 
\begin{equation}
\frac{{\rm d}J(\epsilon)}{{\rm d}\epsilon}= \frac{1}{4\pi}\frac{(\alpha+2)N_uV_u}{[\epsilon_{\rm in}(\epsilon_{\rm in}+2)]^\frac{1}{2}}\left[\frac{\epsilon(\epsilon+2)}{\epsilon_{\rm in}(\epsilon_{\rm in}+2)}\right]^{-\frac{1}{2}\alpha} \quad {\rm \frac{particles}{m^2s\,ster}}
\end{equation}
In this expression $\epsilon\to\epsilon/m_ic^2$ is the particle energy normalised to the particle rest energy $m_ic^2$, and $N_u$ is the density of particles with $p\geq p_{\rm in}$ far upstream. It is remarkable that in this special solution with these special boundary conditions the asymptotic distribution of the particles in independent of $\kappa$. The shock plays a role only in providing the shock compression ratio $\rho_{\rm sh}$, viz. offering the particles the possibility of changing border from downstream to upstream and vice versa many times until a finite state is reached. Accounting for the diffusion is, however, possible \citep[cf, e.g,][]{Jones1991}. It changes the power in the spectrum, which now becomes
\begin{equation}
\alpha' = 1+\frac{3V_u}{8\kappa}\frac{\alpha +1}{\alpha}\Delta
\end{equation}
$\Delta$ is the width of the shock transition layer, which had been assumed to be infinitesimally thin before. In the case of accounting for the diffusion it cannot be neglected anymore. However, following our previous discussion, this is still not correct simply because of the limiting extent of the downstream region which would cut off the integration in $x$ at some $x_{\rm max}=\Delta_{\rm d}$, where $\Delta_{\rm d}\gg\Delta$ is the width of the downstream region. \index{diffusion!coefficients} This can be included by adding the exponential cut-off factor derived in the former section which accounts for losses through the distant boundary.

The neglect of the perpendicular diffusion coefficient implies that the distribution has a spatial dependence only along the magnetic field. The upstream spatial dependence is roughly an exponential decay $\propto \exp(-x/\ell_\|)$ away from the shock with e-folding distance $\ell_\|\approx \kappa_{\|u}/V_u$. This distance depends on energy, remembering that the parallel diffusion coefficient $\kappa_\|=v\lambda_{\rm mfp}$ also depends on energy through both the velocity $v$ and the mean free path $\lambda_{\rm mfp}$ in collisionless plasma, the latter being a complicated functional that is determined through the collisionless interaction between particles and wave spectrum. Inclusion of perpendicular diffusion, which presumably is important for acceleration at quasi-perpendicular shocks but, as numerical simulations show, plays a role also in quasi-parallel shocks is not yet contained in this theory. It violates the assumption of one-dimensionality and thus brings with it a severe complication of shock acceleration theory. 

\begin{figure}[t!]
\centerline{\includegraphics[width=0.95\textwidth,clip=]{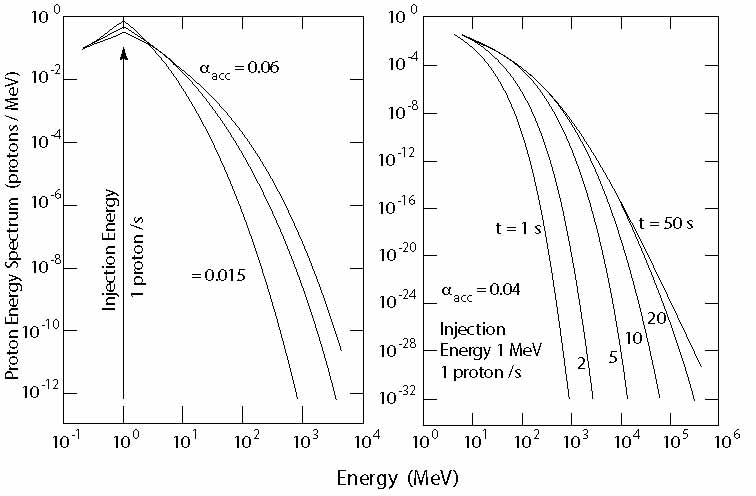}}
\caption[proton spectra]
{\footnotesize Proton energy spectra in first order (stochastic) Fermi acceleration in Alfv\'enic turbulence.  The wave spectrum has  been assumed power law $W_k\sim k^{-s}$. {\it Left}:  Dependence on the acceleration efficiency $\alpha= (V^2/c\lambda_{\rm mfp})\tau_{esc}$ in final state. The escape time has been assumed 1 s.  Protons are continuously injected at energy 1 MeV. The spectrum is no power law but an exponential function. {\it Right}: Time evolution of the spectrum after continuous injection of protons for one given efficiency. The spectrum evolves into harder and harder exponential shape \citep[after][]{Miller1990}.}\label{chap6-fig-millee1}
\end{figure}

\subsubsection{Miller's non-selfconsistent approach}
\noindent Figure\,\ref{chap6-fig-millee1} shows a non-selfconsistent calculation of the particle spectra that are expected to be produced by a power law wave spectrum. Here protons are injected continuously (as assumed by the shock) at energy 1 MeV into the spectrum. Solution of the diffusion equation the produces an exponential spectrum which evolves in time until a stationary state is reached which depends on the acceleration efficiency.

This  not self-consistent  theory has been developed by \cite{Miller1990} in view of application of stochastic (second-order Fermi) acceleration to solar flare \index{acceleration!in solar flare} proton acceleration and electron heating. This approach uses constant escape times and a given wave spectrum and assumes that the wave spectrum decays like a power law $W_k\propto k^{-s}$. The resulting ion spectrum is an exponential, similar to those spectra obtained by \cite{Lee1982} in the quasilinear self-consistent diffusive (first-order Fermi) calculations (see below). However, in this way it becomes possible to check which waves contribute most to the acceleration. The result suggests that it are indeed Alfv\'en waves that play the dominant part in the acceleration through the cyclotron resonance, as was assumed in all the above theories. Magnetosonic waves, which participate in acceleration through Landau resonance, turn out to be much less important. 

In spite of the unrealistic assumption of the power law wave spectrum and the non-selfconsistent treatment the \cite{Miller1991} approach goes one step ahead of the below self-consistent  \cite{Lee1982} quasilinear calculation as it in addition to quasi-linear wave particle interaction accounts for nonlinear wave-particle interactions -- which are also familiar as `nonlinear Landau damping' -- according to the elementary resonance condition $\omega_1-\omega_2=({k}_1-{k}_2)v_\|$, between two Alfv\'en waves $(k_1,\omega_1)$ and $(k_2,\omega_2)$ and low energy protons of parallel speed $v_\|$. This leads to the claim that nonlinear wave-particle interaction does indeed efficiently heat the low energy protons and thus might be capable of lifting thermal ions above the initial energy threshold level for stochastic acceleration. If this is the case, then nonlinear Landau damping could help solving the injection problem explaining how thermal particles can become hot enough in shock interaction to enter the acceleration cycle. \index{process!heating, electron}

The reservation one might have with this theory is that, in contrast to the approach of \cite{Lee1982}, which will be discussed below,  this approach is not self-consistent, assumes continuous injection of protons (or ions) at a given relatively high energy (in Figure\,\ref{chap6-fig-millee1} at the substantial value of kinetic energy of 1 MeV) as well as a fixed power law spectrum of waves. Observation of waves in the foreshock does not necessarily support this latter assumption even though a power law range has been found, as we have discussed above. However, the suggestion of inclusion of nonlinear wave particle interaction into Lee's theory might provide an improvement if the interaction time will be shorter than the time for escape from the narrow downstream region.

\subsection{Lee's self-consistent quasilinear shock acceleration model}\label{chap6-section-lee}
\index{process!diffusive acceleration}\index{process!first-order Fermi, selfconsistent}
\noindent Prior to proceeding to simulations and observational evidence for shock particle acceleration at real shocks in interplanetary space we are going to briefly discuss an analytical selfconsistent model calculation of shock acceleration that has been developed by \cite{Lee1982}. 

\subsubsection{The quasilinear equations of particle and wave dynamics}
\noindent \cite{Lee1982} assumed scattering of ions at circularly polarised Alfv\'en waves, in which case the quasilinear pitch-angle diffusion coefficient becomes
\begin{equation}
D_{\mu\mu}=\frac{\pi\omega_{ci}^2}{2B^2}\frac{1-\mu^2}{v|\mu|}\left.W_k(k_\|,z)\right|_{k_\|=\omega_{ci}/v\mu}, \qquad W_k=W_k^++W_k^-
\end{equation}
where, as usual, the power in the magnetic field fluctuations is given by $\langle |{\bf b}|^2\rangle=\int W_k{\rm d}k$, and the signs $\pm$ on $W_k^\pm$ refer to Alfv\'en waves propagating in the direction $\pm z$ parallel or antiparallel  along the magnetic field ${\bf B}=B{\bf e}_z$, depending on magnetic field direction and wave polarisation being circularly left or right handed. This analytical expression allows to express the parallel diffusion coefficient in terms of wave power
\begin{equation}
\kappa_\|(v,z)=\frac{v^2}{4}\int\limits_{-1}^1\frac{(1-\mu^2)^2}{D_{\mu\mu}}{\rm d}\mu=\frac{m_i^2v^3}{2\pi e^2}\left\{\sum_{\pm} \left[ W_k \left(\frac{\pm\omega_{ci}}{v},z \right) \right]\right\}^{-1}
\end{equation}
The perpendicular diffusion coefficient is modelled according to the classical diffusion model $\kappa_\perp=r_{ci}^2/\tau_{\rm diff}$, and the diffusion time is $\tau_{\rm diff}=\lambda_{\rm mfp}/v$. With $r_{ci}\omega_{ci}=v$ and $\kappa_\|=\frac{1}{3}v\lambda_{\rm mfp}$ this yields $\kappa_\perp\sim v^4\omega_{ci}^{-2}/3\kappa_\|$. This essentially implies that the perpendicular diffusion increases in proportion to the wave power. The diffusive convection equation then reads
\begin{equation}
\nabla_z[\kappa_\|(v,z)\nabla_zF(v,z)]-(\xi_1/a)^2\kappa_\perp(v,z)F(v,z)+VF(v,z)=0
\end{equation}
In this form the diffusion equation takes account of a radius $a$ of a magnetic flux tube from where the particles can freely escape by convection perpendicular to the magnetic field. This had been introduced by \cite{Eichler1981} to account for the free escape boundary. In the case of the bow shock this radius would amount to the radius of the bow shock assuming that the entire bow shock is embedded into this flux tube or radius $a\sim 15\,{\rm R_E}$. In such a case $r/a\ll 1$ and the original distribution function has been expanded in cyclindrical symmetry with respect to Bessel functions of which only the zeroeth order function $J_0(\xi_1r/a)$ has been taken into account, and $\xi_1\approx 2.4$ is the first zero of the $J_0$. [Note that such an approach does not account for the finite width of the magnetosheath although one could interpret it in such a way assuming that the entire flux tube that is embracing the bow shock is draped around the magnetosphere behind the bow shock. In this case the cross section of the flux tube would become $\sim a\Delta_{\rm d}$, which allowed to approximate $a\to\sqrt{a\Delta_{\rm d}}\sim 5\,{\rm R_E}$.] 

For a self-consistent treatment that is still quasi-linear the above convection-diffusion equation must be accomplished by an equation for the evolution of the wave power in dependence on the presence of the fast diffuse ion component if it is assumed that the diffuse ions are responsible for the excitation and damping of the waves. Such an equation, for Alfv\'en waves under stationary conditions, can be written
as the (truncated) wave kinetic equation
\begin{equation}
-(V\mp V_A)\nabla_z W_k^\pm =2\gamma_\pm^AW_k^\pm
\end{equation}
where only convective losses and linear excitation of waves with growth rate $\gamma_\pm^A$ have been retained. The boundary condition on this equation is simply $\lim_{z\to\infty}W_k^\pm=W_{0k}^\pm$ is that at infinity the wave spectrum is the interplanetary wave spectrum. With the resonant Alfv\'en growth rate $\gamma_-=- \gamma_+\equiv\gamma_A(k,W_k^\pm,z)$ being a known functional of the distribution function $F(v,z)$ one may redefine the spatial variable $z\to\zeta=\frac{1}{2}\int_0^z{\rm d}z'/\sum_\pm  W_k^\pm$. Then the above wave kinetic equation, on neglecting the small Alfv\'en speed $V_A\ll V$ against $V$, is rewritten into 
\begin{equation}
\frac{\partial W_k^\pm}{\partial\zeta}=\mp\frac{2\gamma_A}{V}\frac{W_k^\pm}{\sum_\pm W_k^\pm} \quad{\rm or}\quad \frac{\partial}{\partial\zeta}(W_k^+-W_k^-)=-\frac{2\gamma_A}{V},\quad \frac{W_k^+W_k^-}{W_{0k}^+W_{0k}^-}=1
\end{equation}
Together with the above kinetic equation and appropriate boundary conditions at the distribution function $F(v,z)$ these equations form a closed quasilinear system for the distribution function $F(v,z)$ and spectral density of the Alfv\'en waves $W_k^\pm$ that has been solved analytically by \cite{Lee1982} for a given set of plasma and shock parameters in view of application of the bow shock but the solution being valid quite generally. 

In this approach the acceleration times for particles to reach a certain energy or momentum $p$ can be formally estimated from \citep{Drury1983}
\begin{equation}
\tau_{\rm acc}(p)\simeq \frac{3}{\Delta V}\int_{p_0}^p\frac{{\rm d}p'}{p'}\left[\frac{\kappa_{u}(p)}{V_{u}}+\frac{\kappa_{d}(p)}{V_d}\right]
\end{equation}
where $\kappa_u,\kappa_d$s are the respective upstream and downstream spatial diffusion coefficients along the shock normal direction (when neglecting the perpendicular diffusion coefficient these are the projections $\kappa_\|\cos\thetabn$). Recently \cite{Giacalone2008} has given upper and lower bounds for this diffusion time (for infinitely expended regions) in terms of the shock density compression ratio $\rho_{sh}= N_d/N_u>1$ as
\begin{equation}
\frac{3\rho_{sh}}{s(\rho_{sh}-1)}\frac{\kappa_u(p)}{V_u^2}<\tau_{\rm acc}(p)<\frac{3\rho_{sh}(1+\rho_{sh})}{s(\rho_{sh}-1)}\frac{\kappa_u(p)}{V_u^2}
\end{equation}
on the assumption that $\kappa_u(p)\propto p^s$ is a power law function of the particle momentum. The upper bound is obtained for constant $\kappa$ which corresponds to no additional downstream turbulence. The lower bound follow for zero downstream diffusion coefficient.

\subsubsection{Self-consistent particle and wave spectra}
\begin{figure}[t!]
\centerline{\includegraphics[width=0.95\textwidth,clip=]{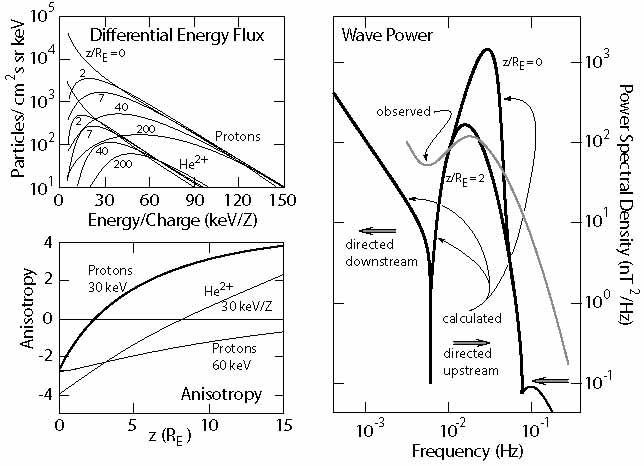}}
\caption[Lee theory]
{\footnotesize Self-consistent particle and wave spectra determined from quasi-linear convection-diffusion theory \citep[after][]{Lee1982}. {\it Left}: Theoretical dependence of differential energy flux  (top panel) on energy per charge number at different upstream distances $z$ from the shock. The sample shock is the Earth bow shock to which all parameters are scaled. Therefore $z$ is measured in Earth radii. It is clear that the particle spectra for both protons and Helium ions (He$^{2+}$) are exponential functions. The spectra drop towards low energies because the theory becomes invalid there, i.e. the diffusion coefficient is increasingly incorrect here. The lower panel shows the particle anisotropy as function of distance from the shock for protons and Helium ions and for two energy channels. {\it Right}: Self-consistent power spectral density  of parallel and antiparallel Alfv\'en waves at two distances $z$ from the bow shock exhibiting a resonant peak at $\sim3$\,mHz. The wave intensity  decreases with distance. These waves are directed upstream (in the plasma frame). At higher and lower frequencies the waves flow downstream. Their high frequency part is the continuation of the low frequency spectrum. There is some similarity between the calculated and measured spectra. }\label{chap6-fig-lee}
\end{figure}

\noindent Figure\,\ref{chap6-fig-lee} given an impression of the power of this self-consistent theory. It shows the calculated differential energy flux for protons and Helium ions as function of energy per charge and the self-consistently generated anisotropy in the particle fluxes in dependence on distance from the shock.  On the right of the figure the self-consistent wave power spectral density is shown for two distances. It is obvious that the interaction between the diffuse ions and the waves generated a large peak of upstream propagating Alfv\'en waves.

The remarkable result of this theory is that the solutions predict ion energy spectra close to the shock which are exponential in energy $\epsilon$ per charge $Ze$ above a certain low energy cut-off with the same e-folding coefficient $\sim 20$\,keV/$Ze$ for all species, which is similar to observation. Moreover, near the shock the ion intensities decrease with distance from the shock with energy dependent scale length which is $\sim 5\,{\rm R_E}$ for 30-keV protons. The solutions also produce an anisotropy in the particle energies which may become quite high, up to $\lesssim 50$\%.   

What concerns the wave spectra so the interplanetary spectrum is strongly modified near the shock by the resonant interaction between waves and diffuse particles. It maximises close to the shock in the resonant frequency range and decreases with distance together with the ion density. It is interesting enough that the waves are found to be polarised essentially linearly suggesting that both polarisations  contribute comparable parts to the wave intensity, which is a little an unexpected result that is explained, however, when referring to the assumed complete isotropy of the distribution with respect to the parallel and anti-parallel propagation along the magnetic field. Thus, there are equal numbers of particles flowing in both directions that resonate with both types of waves. It is important to note that observations do not support this symmetry in the wave spectra.

\subsubsection{Energetic particle diffusion coefficients}\index{diffusion!coefficients, theory}
\noindent In spite of the above noted reservations the theory of \cite{Lee1982} represented a substantial progress in the analytical approach to the self-consistent particle acceleration problem. Apart from the simplifications which it has to make in order to make the problem treatable, its weakest point is the purely known diffusion coefficient. Diffusion coefficients have been calculated under the assumption of quasi-linear theory \citep[first introduced by][]{Jokipii1966} under the assumption of a random walk model of magnetic fields. These models have been shown to be too crude for describing what is observed both in space and in numerical simulations \citep{Giacalone1999}. In this approach the diffusion coefficient is anisotropic {see Eq. \ref{chap6-eq-difftens}) and contains three components, $\kappa_\perp, \kappa_\|, \kappa_A$, where $\kappa_A$ is the Alfv\'enic part used before, resulting from scattering in Alfv\'en waves, and the perpendicular diffusion coefficient is assumed to be related to the parallel diffusion coefficient by
\begin{equation}
\kappa_\perp=\kappa_\|\left[1+\left(\frac{\lambda_{\|{\rm mfp}}}{r_{ci}}\right)^2\right]^{-1}
\end{equation}
and perpendicular transport is usually neglected because the mean free path is much larger than the gyroradius. In this expression the poor knowledge of the mean free path is the big problem as, for instance, there is no reason for having the same mean free path for all particles. \cite{Matthaeus2003} therefore went deeper into statistical theory and proposed an integral equation form for the perpendicular diffusion coefficient
\begin{equation}
\kappa_\perp=\frac{av^2}{3B^2}\int\frac{S({\bf k}){\rm d}^3{\bf k}}{k_\perp^2\kappa_\perp+k_\|^2\kappa_\|+v/\lambda_{b\|} +\gamma({\bf k})}
\end{equation}
where $a$ is a constant of proportionality that is determined from simulations, $\lambda_{b\|}$ is the parallel correlation length of the fluctuating magnetic fields, and $S({\bf k})$ is defined through $S({\bf k},t')=S({\bf k})\Gamma({\bf k},t')$, with $\Gamma=\exp[-\gamma({\bf k})t']$, being the spatial correlation of the transverse magnetic fluctuations
\begin{equation}
\langle b_x[{\bf x}(0),0]b_x[{\bf x}(t'),t']\rangle=\int R({\bf y},t')P({\bf y}|t'){\rm d}{\bf y}
\end{equation}
$R({\bf y},t')$ is the two-point, two-time correlation, and $P({\bf y}|t')$ is the probability  density for a particle at time $t'$ being displaced by the amount ${\bf y}$ in transverse direction. Moreover, the two-time parallel-velocity autocorrelation is modelled by the isotropic assumption $\langle v_\| (0)v_\| (t')\rangle=(v^2/3)\exp(-v_{t'}/\lambda_{b\|})$, and it is assumed that $\gamma=0$. With $P$ being a symmetric Gaussian distribution for the particle trajectory (assuming that the displacement is at all times diffusive), the average over the exponential wave factor $\langle\exp[i{\bf k\cdot x}(t')]\rangle$ which is introduced through the magnetic field fluctuations, yields just $\langle\exp[i{\bf k\cdot x}(t')]\rangle=\exp[(-k_\perp^2\kappa_\perp-k_\|^2\kappa_\|)t']$.
\begin{figure}[t!]
\centerline{\includegraphics[width=0.8\textwidth,clip=]{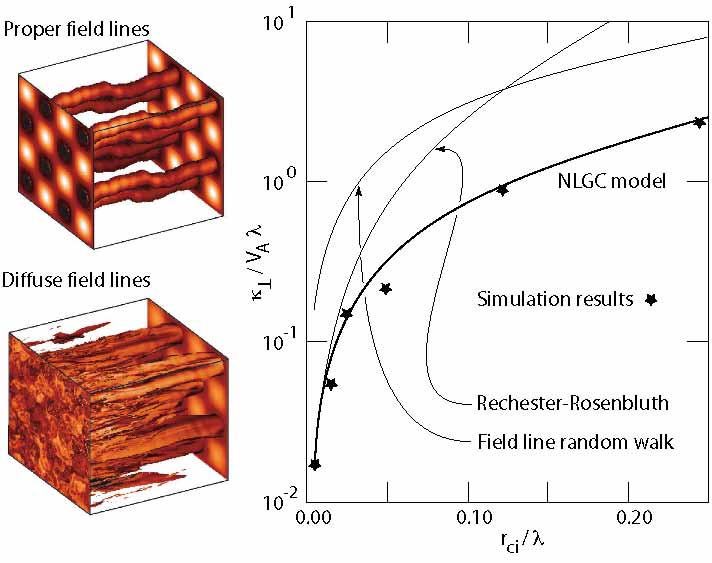}}
\caption[Iapvich]
{\footnotesize Models of the perpendicular spatial diffusion coefficient $\kappa_\perp$ as function of the normalised to $\lambda$ ion gyroradius $r_{ci}$ \citep[after][]{Matthaeus2003}. {\it Left}:  The turbulent magnetic field used in the NLGC diffusion model. The upper part shows a couple of magnetic flux tubes that do not undergo random walk. The field lines remain unbroken. The lower part has 80\% of the magnetic energy in turbulent two-dimensional transverse modes. The magnetic field is highly mixed. This model is used in the numerical NLGC calculations. {\it Right}: The perpendicular diffusion coefficient for several models, the \cite{Rechester1978} model, the ordinary field line random walk (quasi-linear) model, and the NLGC model. The stars show the values obtained from the numerical simulation using the turbulent magnetic field on the left. }\label{chap6-fig-matthdiff}
\end{figure}

This Non-Linear Gyro-Center (NLGC) diffusion model is closer to reality as it takes account not only of the random walk of the fluctuating magnetic field lines but also of the distortion of the particle orbits in this random walk. Diffusion of particles is caused by decorrelation of their orbits from the fluctuating magnetic field. A further development in this theory by using not only Alfv\'enic turbulence but taking into account also the contribution of compressive modes has been presented by \cite{Zank2004}, but the deviations from the NLGC model are very small and can be neglected. 

Figure\,\ref{chap6-fig-matthdiff} shows the used turbulent fluctuation model of the magnetic field and the comparison of the NLGC diffusion coefficient with the simpler models. The agreement between the NLGC diffusion model and the two-dimensional numerical simulation based on the two-dimensional magnetic field turbulence is very good while all other models show severe discrepancies between the simulations of the diffusion and theory. However it is not only this agreement which is of interest. More interesting is that the perpendicular diffusion coefficient is considerably less than quasilinear theory predicts and that the deviation from quasilinear theory become felt already at $r_{ci}\lesssim 0.1\lambda$. Still this model is not yet self-consisten as it ignores the feedback of the particles on the magnetic field turbulence which -- at least to some degree -- is a function of the presence of the energetic particles, as has been suggested by the self-consistent theory of \cite{Lee1982}. Determination of the diffusion coefficient at this stage is merely a necessary though intermediate step.  And event that could still not have been taken into account in an approximate self-conssistnet theory of the kind of the \cite{Lee1982} approach. A complete solution of the problem including parallel and perpendicular diffusion for the relevant waves as well as energy diffusion can thus be expected only from the performance of three-dimensional full-particle PIC simulations of shock formation including particle diffusion and acceleration to high energies. So far no such theory is available yet.
\begin{figure}[t!]
\centerline{\includegraphics[width=0.95\textwidth,clip=]{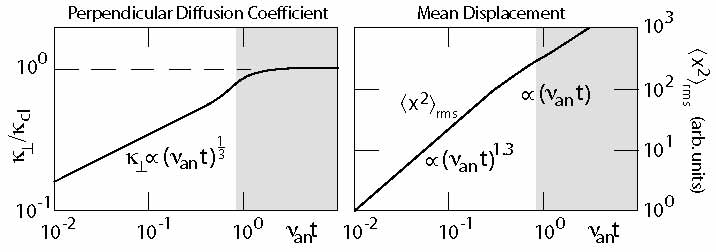}}
\caption[andiffcoeff]
{\footnotesize Evolution of the anomalous perpendicular diffusion coefficient under conditions of L\'evy statistics. {\it Left}: The diffusion coefficient is a power law increasing with time $t$ with anomalous collision frequency $\nu_{\rm an}$. Its maximum slope is $+\frac{1}{3}$. Approaching collision time $\nu_{\rm an}^{-1}$ it merges into the constant classical diffusion coefficient (shaded region). {it Right}: Evolution of the rms perpendicular displacement $\langle x^2\rangle$ as function of time. The maximum slope of the power law is $\sim 1.3$. When classical diffusion takes over the displacement evolves linearly with time.}\label{chap6-fig-andiff}
\end{figure}

\subsubsection{Non-classical diffusion (`super-diffusion')}\index{process!`super-diffusion'}
\noindent The above diffusion coefficients are all based on the assumption of classical diffusion. Since shocks are narrow transitions from one plasma state to another one particle interactions in the vicinity of shocks might be subject to statistics of extremes and not to classical statistics. In this case the diffusion becomes time-dependent as no final state is reached in the process of particle scattering. The probability $P({\bf y}|t')\to P({\bf y}|t';\nu)$ is not anymore Gaussian but has a long tail of power $-\nu$ that extends toward the rare large excursions and long waiting times between the excursions. The theory of such distributions goes back to \cite{Levy1954}. More contemporary developments based on fractal theory have been given by \cite{Shlesinger1993} and \cite{Metzler2000}, and a statistical mechanical argument has been developed by \cite{Treumann2008}. An early application to the derivation of the perpendicular diffusion coefficient \citep{Treumann1997} yields a time dependent  (parallel) mean square particle displacement $\langle z(0)z(t)\rangle =\kappa_\|(d,t;\nu)t$ which leads to a time dependent parallel (or non-magnetised) particle diffusion coefficient 
\begin{equation}\label{chap6-eq-alpha}
\kappa_\|(d,t;\nu)=\kappa_{\|{\rm cl}}[\nu/(\nu-d/2)](\nu_{\rm an} t)^\alpha, \quad {\rm where} \quad \alpha=(4\nu -2d -1)^{-1}
\end{equation}
Here $d$ is the dimension of the system, $\nu_{\rm an}$ is an anomalous collision frequency that is usually nonzero, and $\kappa_{\|\rm cl}$ is the classical diffusivity based on $\nu_{\rm an}$. The condition that $P$ is a positive valued probability distribution is that $\nu-d/2>1$. The diffusion coefficient  then scales as $\kappa_\|\propto T_i\nu_{\rm an}^{\alpha-1}t^\alpha$, which for the scaling with the collision frequency gives a scaling exponent $\frac{2}{3}<|\alpha-1|<1$.  In order to find the scaling of the perpendicular diffusion coefficient one can use the ordinary classical formula $\kappa_\perp=\kappa_\|/(1+\omega_{ci}^2/{\nu_{\rm an}}^2)$ (a more precise theory should use the integral equation for $\kappa_\perp$ derived in the previous section instead of this classical approximation) which gives 
\begin{equation}\label{chap6-eq-kappaperp}
\kappa_\perp= \frac{\nu}{\nu-d/2}\left(\frac{\nu_{\rm an}}{\omega_{ci}}\right)^{\!\!\!2}\,\frac{T_i}{m_i\nu_{\rm an}}(\nu_{\rm an} t)^\alpha
\end{equation}
Under the condition that $\nu_{\rm an}\propto\omega_{ci}$ this yields the scaling $\kappa_\perp\propto \kappa_{\rm B}(\omega_{ci}t)^\alpha$ with $0\leq\alpha<\frac{1}{3}$. It is very important to realise that the entire physics of deviation from purely stochastic diffusion processes is contained in this extraordinarily narrow range of exponents. Very precise determination of this exponent is therefore crucial for elucidating the actual physics of diffusion and hence also diffusive particle acceleration. Below we will discover in numerical simulations of energetic particle diffusion near shocks that the non-classical limit might indeed be realised there.

This kind of transverse diffusion increases with time (see Figure\,\ref{chap6-fig-andiff}) and thus the displacement of the particles grows at a faster rate than any classical transverse displacement and even faster than under Bohm diffusion $\kappa_{\rm B}$. However, if the system would have sufficient time for reaching a final state it would end up in the limit of classical diffusion with linear growing root mean square displacements. This means that under non-classical conditions the system evolves diffusively faster than it would evolve under classical conditions, while given sufficient time it would settle in classical diffusion. A diffusion of this kind is called super-diffusion. In an infinitely extended system it will be realised only temporarily in the initial state for times shorter than the typical classical collision time.

In a spatially limited system like the shock transition the final stationary diffusion state might not be reached, however, since the particles have not sufficient time to undergo classical collisions. In this case the above classical theories require correction for time dependent diffusion, and a stationary state is achieved only when the time dependent diffusion is balanced by losses at the moment when the mean displacement exceeds the size $\Delta_{\rm d}$ of the downstream region. From the definition of $\kappa_\perp=\Delta_{\rm d}^2/\tau_D$ one may estimate the limiting diffusion time $\tau_D$ as 
\begin{equation}
\nu_{\rm an} \tau_D =\left[\frac{\Delta_{\rm d}^2}{r_{ci}^2}\left(1-\frac{d}{2\nu}\right)\right]^\frac{1}{\alpha+1} \Longrightarrow\quad \left(\frac{2}{d}\frac{\Delta^2_{\rm d}}{r_{ci}^2}\right)^\frac{3}{4}<\nu_{\rm an}\tau_D<\frac{\Delta_{\rm d}^2}{r_{ci}^2}
\end{equation}
The left limit holds for the extreme case $\nu-d/2=1$, the right for $\nu\to\infty$ (the latter corresponding to a Gaussian probability distribution). For the bow shock with the downstream region being the magnetosheath of width $\Delta_{\rm D}\sim 2\,{\rm R_E}$ the relative diffusion times will be roughly $\nu_{\rm an}\tau_D\gtrsim 2\times 10^4$. Measuring the limiting diffusion time $\tau_D$ of the energetic particles provides an opportunity to determine the anomalous collision frequency $\nu_{\rm an}$ that governs the diffusive interaction. \index{diffusion!`super-diffusion'}

\subsubsection{Shock simulation of particle diffusion}
\noindent In order to confirm the diffusive nature of the ion acceleration process corresponding to the region downstream of a quasi-perpendicular ($\thetabn=87^\circ$) shock (the Mach  number was ${\cal M}_A=4$), \cite{Scholer2000} performed three-dimensional numerical hybrid simulations measuring the root-mean-square high energy ion displacement $\langle\Delta x^2\rangle$ in the direction perpendicular to the  magnetic field in the downstream region as a function of simulation time. Due to restrictions of computing power they could follow the process just up to roughly $90 \omega_{ci}^{-1}$. 

The result of these simulations is shown in Figure\,\ref{chap6-fig-schodiff} where $\langle\Delta x^2\rangle$ has been plotted as a function of time $t\omega_{ci}$. Initially the displacement performs a large amplitude damped oscillation until the diffusive equilibrium is attained at about $t\omega_{ci}\sim 40$. For later times the particle displacement increases continuously. However, the exponent of the increase is found not to be unity as was expected for classical diffusion. It is rather larger, being 1.17, which is very close to $\frac{7}{6}$ identifying the diffusion process as `super-diffusion'. (Note that, because of the large number $\sim6.3\times10^6$ of simulation macro-particles used of which 525000 have high energies and contribute to the determination of the mean displacement, and because of the high time resolution, the statistical error of the measurement is less than the line width!) \cite{Scholer2000} note this discrepancy while, nevertheless, insist on interpreting their simulations in terms of the classical diffusion picture. 
\begin{figure}[t!]
\centerline{\includegraphics[width=1.0\textwidth,clip=]{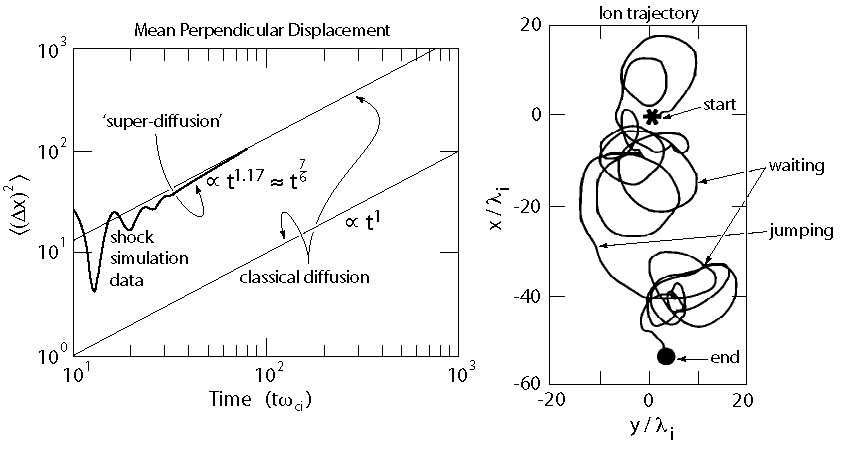}}
\caption[Scholer-diffusion]
{\footnotesize Two dimensional numerical simulation result of the mean downstream perpendicular displacement of quasi-perpendicular ($\thetabn=87^\circ$) supercritical (${\cal M}_A=4$) shock-accelerated ions as function of simulation time \citep[simulation data from][]{Scholer2000}. {\it Left}: The displacement performs an initial damped oscillation before settling in a continuous diffusive increase at about $t\sim 40\omega_{ci}^{-1}$. The further evolution deviates slightly from classical (linear) increase, following a $\langle\Delta x^2\rangle\propto (t\omega_{ci})^{1.17}$ power law.  This is very close to a power of $\frac{7}{6}$, suggesting that the particle diffusion process is in fact `super-diffusive' \citep{Treumann1997}. {\it Right}: Late time trajectory of an arbitrary ion  projected into the plane perpendicular to the mean magnetic field. The ion orbit in the superposition of the ambient and wave magnetic field is not a smooth stochastic trajectory. It consists of waiting (trapped gyrating) parts and parts when the ion suddenly jumps ahead a long distance as is typical for rare extreme event statistics. }\label{chap6-fig-schodiff}
\end{figure}
Above it has been shown that the exponent of the mean displacement in anomalous or `super-diffusion' is  $1+\alpha$ with $\alpha=(4\nu-2d-1)^{-1}$, as given in Eq.\,(\ref{chap6-eq-alpha}) and $0\leq\alpha\leq\frac{1}{3}$. The lower bound corresponds to classical diffusion (the linear increase of the displacement in Figure\,\ref{chap6-fig-schodiff}). In the simulated case $\alpha\approx \frac{1}{6}$ lies clearly in the permitted range of exponents. We may use this value together with the known dimensionality $d=3$ of the simulation in determining the value of the anomalous parameter $\nu=\frac{13}{4}=3.25$. (If using the exact value $\alpha=0.17^{-1}$, we obtain $\nu=3.22$.) This value is sufficiently far above the marginal value of $\nu=\frac{3}{2}$ for the three-dimensional case (or $\nu=1$ for the two-dimensional case). It is, however, also far enough below the classical diffusive limit $\nu\to\infty$ thus identifying the diffusion process as anomalous super-diffusive and non-stochastic (non-Markovian). Since these simulations are completely collision-free, particle diffusion is entirely determined by anomalous processes that are mediated by the self-consistently excited wave spectrum downstream of the shock. 

In view of this reasoning it is instructive to look at the trajectory an arbitrarily chosen fast ion performs in the magnetic fluctuation field. The right part of Figure\,\ref{chap6-fig-schodiff} shows such an orbit in the plane perpendicular to the magnetic field at a late time $100<t\omega_{ci}<160$, when no average diffusive displacement was determined anymore. Clearly, the trajectory is far from being a smooth diffusive particle orbit. It rather consists of a sequence of gyrating sections and sections when the particle breaks out of gyration. When performing the former sections, the particle is `waiting' at its guiding centre location.  In this regime it is probably not interacting with waves. During the break-out sections it suddenly jumps out in order to occupy another waiting position. The beak-out is probably caused by a brief intense wave-particle interaction. The instantaneous location of the chosen particle is progressing only in $x$-direction. Other particles also break out into direction $y$. On a short time scale such a process cannot really be described anymore as being stochastic. 

The non-stochasticity is reflected in the time dependence of the perpendicular diffusion coefficient. However, as we have noted above, this time dependence will last only as long as the time remains to be shorter than the collision time, when the diffusion coefficient will assume its constant classical value. This reasoning implies that {\it the actual diffusion in the collisionless regime is very slow}. This has been concluded (with surprise, because referring to classical diffusion) by \cite{Scholer2000} from the simulations. The diffusion is in fact much less than classical diffusion even though the increase of the displacement is substantially faster than under classical diffusive conditions. Using classical diffusion in diffusive acceleration grossly overestimates the diffusive effect. The term `super-diffusion' does not apply to the strength of diffusion but just to its time-dependence. 
\begin{figure}[t!]
\centerline{\includegraphics[width=0.95\textwidth,clip=]{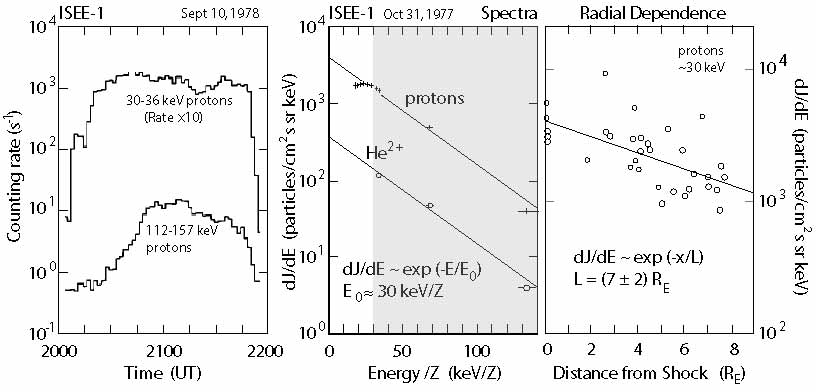}}
\caption[Iapvich]
{\footnotesize Energetic ion spectra upstream of the Earth's bow shock \citep[after][]{Ipavich1981a,Ipavich1981b}. {\it Left}:  The count rate in the upstream region of the bow shock for protons in the channels of 30-36 keV and 112-157 keV. Note the delay of $\sim$\,40 min between the rises of low and high energy fluxes. {\it Centre}: The energy per charge number spectra decay exponentially (shaded region) with same energy-per charge number scale factor for all ions, suggesting energy diffusion. {it Right}: Spectral variation with shock distance. The intensity of energetic particles decays exponentially with distance from shock, which is a strong argument for the shock source and for a diffusive process. }\label{chap6-fig-ipav}
\end{figure}

\section{Observations}
\noindent Observation of energetic ions around shocks has a long history starting in the late sixties of the past century \citep[for a relatively recent review of inner heliospheric observations see][]{Reames1999}. Since the {\footnotesize ISEE} spacecraft it became possible to determine both the spectra of energetic particles and their dependence on distance from the shock \citep[in this case again the Earth's bow shock; for a review of energetic ions in the foreshock see][]{Fuselier2005}. Figure\,\ref{chap6-fig-ipav} shows examples of those observations upstream of the shock. In the left part the count rates for protons in two energy ranges $30-36$\,keV and $112-157$\,keV have been plotted against time when the spacecraft was in the upstream foreshock region observing an energetic event that was connected with the shock \citep{Ipavich1981a,Ipavich1981b}. For two hours the shock injected high fluxes of energetic protons into the upstream region. It is interesting to note that the higher energies occur later. 

Even though it is not known whether this is a purely spatial or a temporal effect, the delay in the energetic particles points on the action of an energy filter which naturally would be given by energy diffusion. The central panel shows the energy spectrum of energetic particles, this time protons and helium nuclei ($\alpha$ particles) plotted per charge number. Note that the plot is half-logarithmic. Hence, in both cases the spectra are exponentials, exhibiting the same scale $\sim 30$\,kev/Z in unit of energy per charge number Z, again reflecting the energy diffusion process. The panel on the right shows the spatial dependence of the differential energy flux for protons of 30\,keV energy. Though the scatter is substantial (the correlation coefficient is 0.62), which is due to the time variability of the fluxes, in the mean the dependence of flux on distance from the shock is described quite well by an exponential decay with shock distance, implying that the energetic particles undergo a spatial diffusion process away from the shock. Hence, the shock is the source of the energetic particles.  
\begin{figure}[t!]
\centerline{\includegraphics[width=0.95\textwidth,clip=]{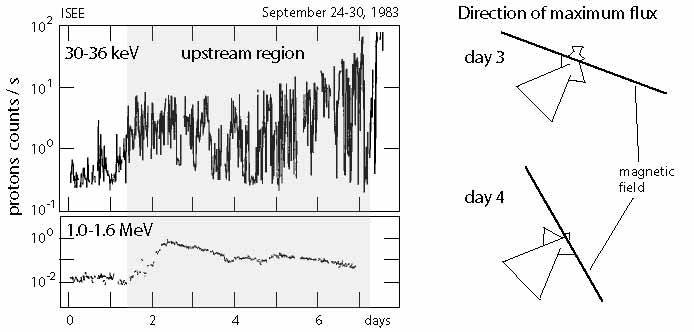}}
\caption[Lee theory]
{\footnotesize ISEE observations of energetic ions upstream of the Earth's bow shock \citep[after][]{Terasawa1985}. {\it Left}:  The count rate in the upstream region of the bow shock (shaded) for protons in the channels of 30-36 keV and 1-1.6 MeV. The enhanced fluxes in the upstream region are well expressed. Moreover, on this long time scale the fluxes are highly variable in time in the lower energy channel. At high energies the time resolution is considerably less, so the variability is suppressed stronger. {\it Right}: The direction of arrival of the maximum 30-36 keV proton flux as seen in the spacecraft frame and relative to the instantaneous direction of the magnetic field for two occasions, day 3 and day 4 of ISEE observations. The counts are anisotropic, preferably distributed perpendicular to the magnetic field.}\label{chap6-fig-tera}
\end{figure}

\cite{Terasawa1985} used {\footnotesize ISEE-3} measurements in order to determine the dependence of the upstream energetic ion fluxes on the direction of the magnetic field. Figure\,\ref{chap6-fig-tera} shows these observations over a time period of seven days. There is a high variability of the energetic proton fluxes in the $30-36$\,keV channel that can be traced back to the variation in the magnetic field direction. On the right of this figure two cases of direction measurements of the fluxes are given. Clearly the maximum flux is directed from perpendicular to the interplanetary magnetic field. Thus the fluxes are highly anisotropic which is partially affected by the particles being convected in this direction by the flow. At higher energies $> 1$\,MeV there is less variability in the proton fluxes which may be due to larger isotropy of the high energy particles while being affected by the lesser resolution. \cite{Terasawa1985} also conclude that the main acceleration of the protons occurs at the shock nose part of the upstream region which is probable as the shock is strongest in this part.
\begin{figure}[t!]
\centerline{\includegraphics[width=0.90\textwidth,clip=]{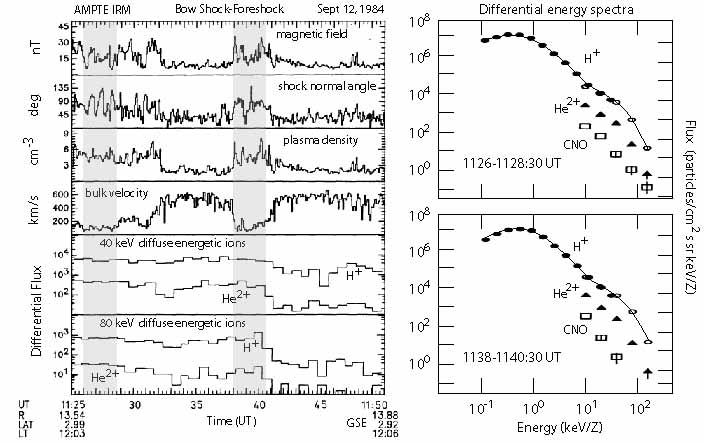}}
\caption[Iapvich]
{\footnotesize Energetic AMPTE IRM ion observations and spectra in the downstream region of the Earth's bow shock \citep[data taken from][]{Ellison1990}. This time interval was chosen because the spacecraft was at the nose of the bow shock and the interplanetary magnetic field was nearly radial providing about stationary quasi-parallel shock conditions. {\it Left}: Plasma and high energy particle data (the lower two panels showing proton and Helium ion fluxes at 40 keV and 80 keV). {\it Right}: Differential energy spectra obtained during the shaded time intervals on the left for protons, helium ions (triangles) and CNO (Z=6; squares). Full and open circles are from two different instruments. A high energy tail evolves on the distribution function at $>10$ keV/Z. Within the measuring error the distributions are identical for all ions.}\label{chap6-fig-hepbs}
\end{figure}

These bow shock observations have in the past been confirmed by other spacecraft \citep[see, e.g., the {\footnotesize AMPTE IRM} observations in Figure\,\ref{chap6-fig-hepbs} reported and successfully modelled by Monte Carlo test particle simulations by][]{Ellison1990}. One might argue that these exponential spectra would be a peculiarity of the conditions at the bow shock, therefore. However, energetic particle spectra downstream of interplanetary travelling shocks \index{shocks!interplanetary} do also decay exponentially with energy. This is shown in most recent observations by the {\footnotesize ACE} spacecraft (see Figure\,\ref{chap6-fig-ace}) and has been confirmed for the same events by {\footnotesize GEOTAIL} near Earth  \citep{Giacalone2008}. Within the certainty of the measurement the slope on both spacecraft was the same. The intensity difference may have had many reasons, being affected by the travel time between the two spacecraft ({\footnotesize ACE} was more than 1 Million km upstream in sunward direction of {\footnotesize GEOTAIL}) \index{spacecraft!GEOTAIL} and the geometry of the shock and environment. Obviously the acceleration physics is similar at bow shocks and at travelling shocks. Both kinds of shocks accelerated ions to higher energies when being exposed to scattering centres to both sides of the shock. Such scattering centres are abundant in the foreshock region and downstream of the shock due to the selfconsistently excited wave spectra  that are present in the close vicinity of the shock. The intensity of these waves decays at large distances from the shock and the scattering process will then rely solely on the general turbulence present in the stream. 

The selfconsistent theory and observations near Earth demonstrate that the accelerated particle spectra are exponential. Hence, the acceleration mechanism generating these particles is of the kind of the first-order Fermi mechanism (or diffusive acceleration). This is in contrast to  the power spectra observed at higher energies in cosmic rays. This discrepancy has not yet been resolved while there are indications that the higher energy particles do in fact behave more like test particles and do not become so much involved into the selfconsistent shock formation/acceleration process. Moreover, at higher energies the relativistic nature of shocks comes into play thereby completely changing the picture one naively has of the process of acceleration. Shock acceleration is then closer to the original second-order Fermi process with the result that the spectrum becomes power law over some extended range in energy the extension of which depends on the width of the acceleration region and the loss mechanism. At the end of this range an exponential cut-off will  limit  the spectrum. The cosmic ray spectrum of Figure\,\ref{chap6-fig-CR} suggests however that this cut-off for the energetic cosmic rays lies at extraordinarily high energies. Nothing similar applies to the shocks in the heliosphere. It thus seems that these are in the majority located in the regime of first order Fermi acceleration and do not generate a clear power law. 
\begin{figure}[t!]
\centerline{\includegraphics[width=0.50\textwidth,clip=]{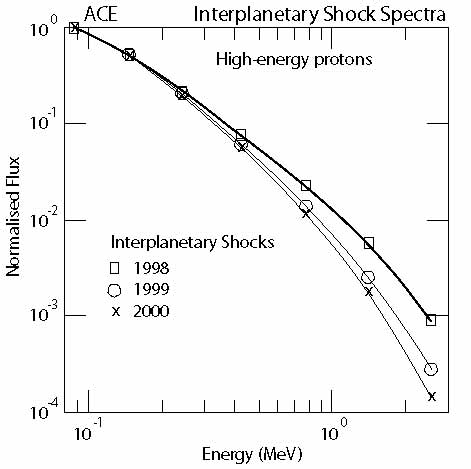}}
\caption[Iapvich]
{\footnotesize Energetic post-shock ion spectra at three travelling interplanetary shocks observed by the ACE spacecraft far upstream (about 1-2 Mill. km)  in the solar wind \citep[after][]{Giacalone2008}. Nearly identical spectral slopes though of varied intensity were observed for the same shocks during their passage near Earth by GEOTAIL. The spectral shape of the accelerated protons at the interplanetary shocks is as well exponential.}\label{chap6-fig-ace}
\end{figure}

\section{The Injection Problem}\noindent
One of the most tantalising problems in shock acceleration physics is the origin of the energetic  seed ion population that is required for starting the acceleration cycle. How are the relatively large numbers of ions generated by the shock that have energies above the injection threshold? In anticipating the result of the following discussion of a few models that have been proposed we note that there is no solution in sight yet at this instant.  \index{process!shock injection}\index{shocks!injection problem} Since the problem is extraordinarily complex it can be investigated only on the way of numerical simulation studies and, at its best, presumably only with the help of self-consistent full particle PIC simulations adopting real ion to electron mass ratios.

There are two ways for treating the complicated process of shock particle acceleration by numerical simulation: test particle simulations and selfconsistent simulations of shock formation. The classical way has for long time been performing non-selfconsistent test particle simulations. Because the diffusive process of particle acceleration is a slow process, selfconsistent simulations require long times and large simulation boxes. Accounting for both has partly become possible only in most recent times. In the following we mention test particle simulations only in passing before turning to a deeper rooted recollection of recent results on selfconsistent mostly hybrid and low-dimensional simulations, even though test particle simulations may come close to reality in extraordinarily large systems. Such systems are available in astrophysics for the small fractions of accelerated particles in Cosmic Rays. The space physics environment, on the other hand, requires selfconsistency because the shock scales are finite, energies of shock-accelerated particles are comparably low, and the numbers of accelerated particles still substantial. Under such conditions the interaction between the accelerated particles and the plasma cannot be neglected.

\subsection{Ion shock surfing} \index{process!shock surfing, ions}\index{shocks!ion surfing}
\noindent Since quasi-perpendicular shocks with their shock potential reflect low energy ions out of the upstream thermal ion population one may expect that this ion reflection causes ion acceleration due to multiple reflection \citep{Sagdeev1966,Sagdeev1973,Ohsawa1985a,Ohsawa1985b}.
This shock electric potential field can be estimated from the electron momentum equation, neglecting the motional electric field,
\begin{equation}
e\phi(x)\approx \int_{-\infty}^x{\rm d}x\frac{1}{N}\frac{\partial}{\partial x}\left(\frac{B^2}{2\mu_0}+P_e\right)
\end{equation}
Making use of the constancy of the total pressure $[B_2^2/2\mu_0+P_{e2}+m_iV_2^2]=0$ and neglecting the electron pressure in first approximation (see the above discussion) this can be rewritten, yielding
\begin{equation}
e\phi(x)\approx \frac{B_1}{\mu_0N_1}\int_{-\infty}^x{\rm d}x\frac{\partial B}{\partial x} = \frac{1}{{\cal M}_A^2}\frac{\Delta B(x)}{B_1}\,m_iV_1^2
\end{equation}
where $\Delta B(x)=B_2(x)-B_1$ is the difference between the downstream and upstream magnetic fields taken at position $x$. For ions this potential implies that ions with energy $\frac{1}{2}m_iv_x^2<e\phi(x)$ will be reflected at position $x$ because the potential is positive. As discussed in the chapter on quasi-perpendicular shocks these reflected ions are partially responsible for the quasi-periodic reformation of quasi-perpendicular shocks. When gyrating back into the upstream plasma they form the shock foot, retard the inflow, cause a foot current, and most important experience the motional electric field in which they become accelerated along the shock surface until gaining sufficient energy to pass the reflecting shock potential. Clearly, their acceleration is on the expense of the upstream motional electric field, and consequently the inflow is slightly retarded in this acceleration process. 

The maximum energy gain of the accelerated ions in this multiple upstream reflection before passing the shock, which we can call `ion surfing', can be estimated from the balance between the Lorentz   and the electric forces on the ion, yielding the simple upper limit for the tangential to the shock ion velocity $v_{y,{\rm max}}\sim {E_x}/{B_z}$, where $E_x$ is the cross shock reflecting electric field based on the above potential. Combination with the above result then produces the maximum energy \index{energy!gain,ion maximum}
\begin{equation}
{\cal E}_{i,{\rm max}}\sim \frac{m_i}{2}v_{y,{\rm max}}^2\sim \frac{m_i}{2}\left(\frac{1}{{\cal M}_A^2}\frac{m_iV_1^2}{eB_1\Delta_s}\frac{\Delta B}{B_1}\right)^2
\end{equation}
an ion can gain in this multiple reflection process, where $\Delta_s$ is the thickness of the cross shock ramp electric field region. Hence the maximum energy gain depends crucially on the narrowness of this region. Since the shock thickness is not well known and the effect of the overshoot field is difficult to estimate, one assumes that in the overshoot region the magnetic field can be described approximately by a solitary structure yielding an estimate $\Delta B/B_1\sim 2({\cal M}_A-1)$ for the maximum overshoot amplitude. With this estimate the maximum energy gain becomes very large 
\begin{equation}
{\cal E}_{i{\rm max}}\sim 2m_iV_A^2{\cal M}_A^2(m_i/m_e)=2m_iV_1^2(m_i/m_e)
\end{equation}
In fact, in the solar wind the streaming energy of ions if of the order of 1 keV, implying that the maximum energy an ion could gain in surfing along the bow shock amounts to the order of ${\cal E}_{i{\rm max}}\sim 7.5\,{\rm MeV}$ quite high. This energy would be sufficient for entering the Fermi acceleration cycle, while the observations suggest a steep cut off of the proton spectrum at the bow shock above a few 100 keV.  This could be due to the curvature of the shock that does not allow a proton to surf for sufficiently long time. But the question remains unresolved whether or not the bow shock and thus any shock can by itself accelerate ions by surfing to exceed the energy threshold of the Fermi cycle.

\subsection{Test particle simulations}
\noindent Test particle simulations assume a shock that is given for granted and is not (or very little) affected by the accelerated particles. This holds as long as the density of the high energy particles remains much less than the ambient plasma density in the flow. The shock can be prescribed by the Rankine-Hugoniot conditions, assuming a given shock compression ratio (thus implicitly assuming a Mach number), and the particles are subject to some stochastic scattering mechanism in a given turbulent wave spectrum after having been injected at some high enough energy such that the shock can be safely considered infinitesimally thin with no substructure. The wave spectrum is usually modelled by a power law
$W_k\propto [1+(|k|\lambda_{corr})^\varsigma]^{-1}$ of purely isotropic magnetic fluctuations. Here $k$ is the wave number, $\lambda_{corr}$ the wave correlation length, and $\varsigma$ the wave spectral index. The wave correlation length is modelled as $\lambda_{corr}\propto V_1/\omega_{ci}$ proportional to the gyroradius of the flow. Taking the constant of proportionality sufficiently large, being of the order of $10^3-10^4$ ensures that the injected ions perform many gyrocircles before getting out of contact with the waves. In such a spectrum of waves particles `diffuse' across the magnetic field by following the meandering magnetic field in the magnetic fluctuations.

The part of the distribution below the injection threshold is not affected by the presence of the test particles, and to good approximation it is assumed that the Rankine-Hugoniot jump relations remain intact during the acceleration process. Ions are injected at energy sufficiently high above the injection threshold and are assumed to be reflected back and forth across the the infinitesimally thin shock in the wave turbulence until moving out to higher energies and creating a power law tail in the high-energy part of the ion distribution. This tail is cut off exponentially at the loss boundary as has been discussed before. We repeat that the particles are accelerated only in the upstream scattering by the upstream motional electric field ${\bf E = - V\times B}$. In the downstream scattering process the particles are overtaking the waves and thus loose energy. 
\begin{figure}[t!]
\centerline{\includegraphics[width=0.95\textwidth,clip=]{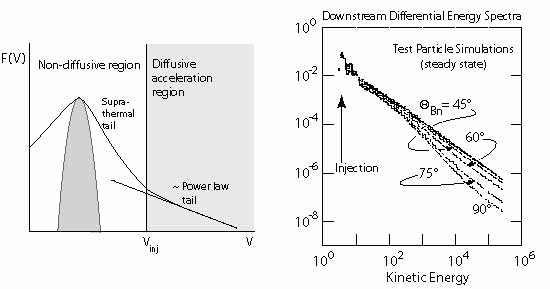}}
\caption[Giaca]
{\footnotesize Test particle simulations of shock acceleration injecting protons into a pre-existing shock at different shock normal angles $\thetabn$ \citep[after][]{Giacalone2005a}. {\it Left}:  Schematic of shock acceleration. Below injection velocity $V_{\rm inj}$ it is assumed that the selfconsistent shock formation process produces a nonthermal tail. Sufficiently energetic ions will be capable of crossing the injection velocity threshold at the lower bound for the diffusive shock acceleration cycle. Above injection velocity a power law tail evolves due to the acceleration process. Test particle simulations are performed in this regime injecting test particle ions at an energy that is sufficiently far above the injection threshold. {\it Right}: Test particle spectra obtained for different $\thetabn$. Energy is in units of the upstream flow kinetic energy. The hardest spectra with clearest power law tail are obtained for small $\thetabn$, but the dependence is weak.} \label{chap6-fig-testsim}
\end{figure}\index{acceleration!test particle}\index{simulations!test particle}

Figure\,\ref{chap6-fig-testsim} on its right shows an example of the power law tail spectra obtained in such test particle simulations \citep[performed by][]{Giacalone2005a}. The spectra are given in units of the upstream flow energy. The injection energy is taken as 1 MeV, and the final steady state spectra are given for different shock normal angles $\thetabn$. One notes that the flattest spectrum is obtained for small $\thetabn$ in the quasi-parallel shock. The spectra show that irrespective of the shock normal angle particles are accelerated assuming a final steady state spectrum. This spectrum becomes power law above roughly $10^3$-times the upstream flow energy for the quasi-parallel and $10^4$-times for the perpendicular shocks. Above about this energy the slopes of the spectra are comparable for both quasi-perpendicular and quasi-parallel shocks. The difference is only in the flux with the quasi-parallel shock generated flux at given energy about one order of magnitude larger than the quasi-perpendicular shock generated flux. Nevertheless, time dependent test particle simulations of the same kind show that even though the fluxes are low the quasi-perpendicular shock accelerates particles to higher energy at a given time than the quasi-parallel shock. In other words, quasi-perpendicular shocks show a higher acceleration rate than quasi-parallel shocks. 

\cite{Giacalone2005a} also investigated the dependence of the acceleration efficiency on the strength of the magnetic fluctuations. He found that with decreasing strength the fluxes are lower but the slope of the distribution remained about the same. On the other hand, the presence of very long scale waves turns out to be important. The longer the wave and correlation lengths the more efficient is the acceleration; this holds in particular at quasi-parallel shocks while quasi-perpedicular shocks are relatively insensitive to this variation. This effect is probably caused by the variation in the shock normal angle $\thetabn$ close to the quasi-parallel shock that is related to the presence of long scale waves. Near the shock transition these transverse waves which in the spectrum have the largest amplitudes provide the quasi-parallel shock a less parallel (or more quasi-perpendicular) character thereby enhancing the acceleration efficiency. 

\subsection{Self-consistent shock acceleration simulations}
\noindent While at the largest energies the dependence of the acceleration is rather insensitive to the character of the shock, self-consistent numerical simulations are the only means of receiving theoretical information about the acceleration efficiency and the very mechanism of acceleration at the shock. The lower energies are very sensitive to the nature and structure of the shock and, vice versa, the shock depends strongly on the properties of the low energy particles as long as their population forms part of the main flow.This is the case for all observed deformed particle distributions in the heliosphere where  the energetic tails make up a non-negligible fraction of the main distribution. Their generation requires performing particle PIC simulation, hybrid if just the ions are the primary subject of interest, or full particle PICs if the dynamics and acceleration of electrons and their modification of the ion dynamics are in question. In both cases, for investigating reflection and acceleration, the simulations should be higher dimensional, firstly because the particle motion consists of the superposition of the two-dimensional gyromotion and the parallel particle dynamics, secondly because of the two-dimensionality of the diffusion process occupies a central place in the acceleration of particles and their escape from the shock into space. 

In the present section we review the available more recent numerical simulations of particle acceleration at shocks restricting ourselves to the non-relativistic subset of these simulations. Our main purposes of doing so is to receive information about the theoretical possibilities of shock-injection of particles into the Fermi acceleration process. We, moreover, treat the acceleration or heating of ions and electrons separately paying attention first to ion acceleration and injection.

\subsubsection{Downstream ion heating - leakage from downstream to upstream}\noindent Quite early \cite{Tanaka1983} suggested a mechanism for ion heating and injection from downstream by wave scattering. The idea of this mechanism is triggered by the observations of \cite{Sckopke1983} that shock reflected ions from a quasi-perpendicular shock surface after having been accelerated in the upstream convection electric field (which, we remind, is geometrically parallel to the ion gyration in the quasi-perpendicular shock foot; note also that is these ions that contribute to quasi-perpendicular shock reformation!) gain high enough energy to overcome the shock ramp and enter the downstream region. This may take several cycles of ion reflection in gyration in the quasi-perpendicular shock foot. 

When having passed the shock ramp and overshoot and entered the downstream region, these ions possess a large excess thermal energy in the perpendicular motion, $T_{i\perp}\simeq  (3-5)T_{i\|}$, which is sufficient to excite unstable electromagnetic ion-cyclotron waves in the downstream region just behind the shock. Observations do indeed show that these waves form part of the downstream electromagnetic spectrum \citep[][]{Narita2006a,Narita2008}. Quasilinear wave particle interaction between the ion-cyclotron waves and the energetic ions scatters the ions in pitch-angle thereby approaching isotropy. Some of the ions then will escape upstream along the magnetic field adding to the upstream energetic ion component. \index{acceleration!leakage from downstream}

The simulations of \cite{Tanaka1983} suggest that the escape is preferentially for $\thetabn\sim 45^\circ$, and the ions have roughly parallel energy two times the upstream flow energy. This is still much too small for entering the acceleration cycle. However, these simulations assume a plane shock at fixed $\thetabn$. In a curved shock like the bow shock, the ions will escape upstream at the quasi-parallel part of the shock where escape is easier. Arriving upstream they will again become accelerated in the perpendicular direction by the convection electric field up to a perpendicular energy roughly a factor of two higher. Passing the shock another time the cycle will be repeated with self-scattering in pitch angle until some of the particles may reach energy high enough for entering the first-order Fermi acceleration cycle. This might be one possibility for feeding first order Fermi acceleration. However, there is only qualitative information available and no selfconsistent simulation exists so far which could support this claim. Moreover, this process is slow and probably not fast and efficient enough for ion injection. A sufficiently intense  downstream wave spectrum is clearly required for pushing ions back across the shock barrier upstream by scattering. But this scattering itself is insufficient of arming the ions with the required jump in energy to overcome the injection threshold even though the ions when re-entering  the shock from downstream feel the shock potential which further accelerates them in the upstream direction helping them to pass the shock back upstream against the flow.

\subsubsection{Hybrid simulations}
\noindent One-dimensional and two-dimensional hybrid simulations of self-consistent shock formation with the focus on particle acceleration have been performed for the past roughly two decades with increasing resolution \citep[][and others]{Burgess1987,Scholer1990,Kucharek1991,Scholer1999,Giacalone2004,Giacalone2005b}.
\begin{figure}[t!]
\centerline{\includegraphics[width=1.0\textwidth,clip=]{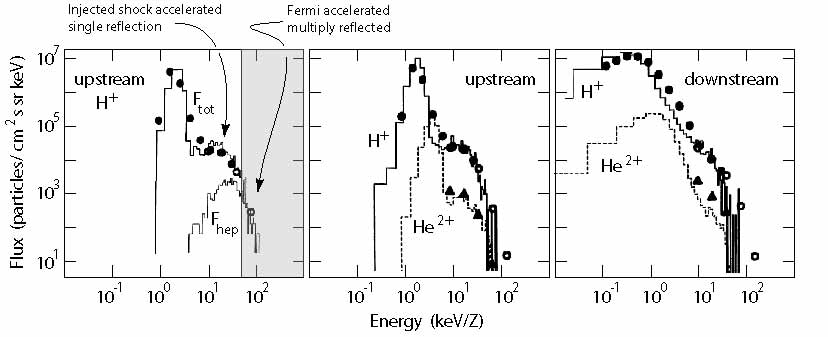}}
\caption[injection]
{\footnotesize One-dimensional hybrid simulation with ion splitting showing the acceleration of ions up to the injection boundary for multiple reflection (first order Fermi) acceleration \citep[after][]{Scholer1992}. {\it Left}:  The simulated upstream proton spectrum compared with the observations of \cite{Ellison1990}. The total proton flux $F_{tot}$ agrees satisfactorily well with the observations. The distribution exhibits the Maxwellian upstream flow and a broad shock-accelerated exponential bump at energies $>10$ keV, typical for shock acceleration. The distribution $F_{hep}$ contains the population that has been accelerated by the shock but is scattered back towards the shock thus becoming multiply scattered. Above $\sim 50$ keV it makes up the total distribution identifying the shaded region as the Fermi acceleration domain. {\it Centre}: Similar upstream simulation for protons and Helium nuclei. {\it Right}: The downstream distribution for the two particle components fitted to the observations. Here the data points are the same as in Figure\,\ref{chap6-fig-hepbs}. The agreement between observation and simulation is reasonable.} \label{chap6-fig-qpinj}
\end{figure}\index{shocks!acceleration, particle injection}\index{acceleration!shock particle injection}

\cite{Scholer1992} tried to reconstruct the {\footnotesize AMPTE IRM} spectra observed by \cite{Ellison1990} with the help of self-consistent hybrid simulations of acceleration at a quasi-parallel $\thetabn=30^\circ$ shock in order to check the validity of the test particle approach used by those authors and to infer about the possible self-consistent ion injection mechanism. Since the amount of self-consistently produced shock-accelerated ions in the simulations was very small they applied the trick of splitting each reflected ion into 30 parts thereby generating a factor of 30 more ions than the simulation produced while not changing the total charge and mass of the ion component. This splitting was applied only to the reflected ions once an ion was recognised as being reflected. Otherwise the simulation was one-dimensional and computer-time limited. There were a number of further restrictions that were related to the upstream boundary of particle escape which was assumed fixed in space such that the distance shock-escape boundary continuously shrunk during the simulation, and a finite resistivity had to be assumed making the simulations not fully collisionless.    

Figure\,\ref{chap6-fig-qpinj} shows the results of these simulations for a pure proton run (left) and a run including protons and helium nuclei /centre and right) compared to the observations of \cite{Ellison1990}. The fits of the simulated and observed data are satisfactory for both the upstream and downstream particle energy fluxes. The spectra obtained are exponentials and no power laws. Similar simulations in one dimension in a larger domain but assuming the presence of a given upstream wave turbulence spectrum \citep{Scholer1999a} and with better resolution yield also exponentials for both particle species, protons and helium nuclei.

The important physics is, however, contained in the left panel which shows the results for a pure proton hybrid simulation with a large number of particles. The two flux curves in this panel belong to the total particle flux upstream which consists of the Maxwellian flow and a shock accelerated exponential bump on the tail of the distribution which is completely self-consistently produced by the shock. The simulations do not tell in which way these protons have been accelerated. However, since in the simulation the accelerated particles had been given a flag, it was possible to plot the distribution of those particles separately which had been scattered back towards the shock from upstream. These particles have the distribution $F_{hep}$ and obviously will, after reaching the shock, have another chance to be  reflected again and further accelerated. In fact, above $\sim50$ keV (in these simulations which have been tailored for  bow shock conditions) the total high energy flux is completely built up of these multiply reflected particles. Hence this is the domain of first order Fermi acceleration where the particles become accelerated.

We may therefore conclude that the quasi-parallel shock is capable of accelerating ions into an exponential energetic bump. Part of the flux in this energetic bump becomes reflected back to the shock and has the chance of becoming accelerated in the Fermi mechanism. Parallel shocks are thus capable of providing the pre-accelerated particle component that becomes the seed population for shock Fermi acceleration. Concluding from these simulations the quasi-parallel shock itself is capable of generating a see population for further first order Fermi acceleration. It does not need to be fed by particles that are reflected at the quasi-perpendicular part of the curved shock surface. Further hybrid simulations by \cite{Scholer1998} support this conclusion by showing that the quasi-parallel shock-generated ion spectrum extends into the energy range of a few ten times the upstream ram flow energy. Thus the quasi-parallel shock generates its own seed population and does not need any help by a quasi-perpendicular shock. This is a conclusion of vital importance in application to astrophysics. 

However, the very mechanism of how the parallel shock is able to accelerate the particles in order to inject them into the Fermi cycle remains an open problem. It is not answered by these simulations except that they clearly show that the energetic particle component is not a population that is leaking out from the downstream region into the upstream region as proposed in the mechanism of \cite{Tanaka1983}. The seed particles are a  part of the thermal upstream distribution that has become accelerated by the quasi-parallel shock in a first step followed by a second step in the acceleration chain. 

Similar simulations in one dimension but a much larger spatial simulation domain $x\geq 28000\lambda_i$  and simulation times $t\omega_{ci}=4000$ have been performed by \cite{Giacalone2004} for a parallel shock of Mach number ${\cal M}_A=6.4$. Like in the simulations of \cite{Scholer1992}, the differential energy flux spectrum exhibits a shoulder at high energies. However in this simulation, because of the large simulation domain and long simulation times, these shoulders extend out to energies up to 200 times the initial upstream plasma ram energy (in application to the solar wind an acceleration up to 200 keV).

\cite{Giacalone2005b} has recently investigated the complementary problem of acceleration at a nearly perpendicular shock in his half-selfconsistent numerical simulations. Here the `half' refers to the fact that a turbulent magnetic wave spectrum is superimposed on the magnetic field with spectral index $\varsigma=\frac{5}{3}$ simulating Kolmogorov wave turbulence.  His results are shown in Figure\,\ref{chap6-fig-hepsim}. The peculiarities of this simulation are the two-dimensionality and the spatial extendedness along the mean magnetic field, which is about parallel to the shock surface. Such a simulation geometry allows including long wavelength magnetic perturbations.  In the case of field line meandering this should enhance any perpendicular particle diffusion. The assumed Mach number is ${\cal M}_A=4$. As in hybrid simulations usual a small anomalous resistivity is included. The imposed turbulence together with the self-consistent shock dynamics affects the shock surface which becomes wave and rippled. The longest wavelength is of the order of the size of the simulation box which was also assumed to be the turbulent correlation length.
\begin{figure}[t!]
\centerline{\includegraphics[width=1.00\textwidth,clip=]{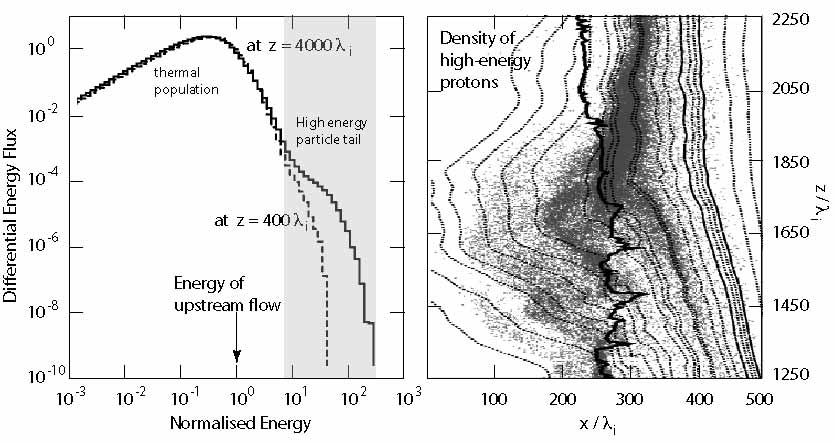}}
\caption[Giaca]
{\footnotesize Two-dimensional self-consistent hybrid simulations of generation of energetic protons in an extended simulation box around a nearly perpendicular shock \citep[after][]{Giacalone2005b}. {\it Left}:  The  spectra at two different locations. The high energy tail that has evolved at the distribution is more expressed as the distant location. {\it Right}: The energetic proton distribution in the ${x,z}$-plane and the field lines around the shock at time $t\omega_{ci}=150$. Plotted are only the protons in the tails of the distribution (shaded on the left at two locations). The shock ramp is the dark line, the field lines are the light lines. Energetic particles concentrate almost exclusively inside the looping magnetic field lines. It seems that here the strongest acceleration takes place.} \label{chap6-fig-hepsim}
\end{figure}\index{simulations!self-consistent hybrid acceleration}

The right part of Figure\,\ref{chap6-fig-hepsim} shows the instant $t\omega_{ci}=150$ in part of the simulation box of length $\Delta z=1000\lambda_i$. The shock surface is located at the dark line undulating between $x=200\lambda_i$ and $x=300\lambda_i$ along $z$. The light lines are the magnetic field lines. The dots are the instantaneous locations of the energetic particles with energy $>10\times \frac{1}{2}m_iV_1^2$, i.e. larger than ten times the upstream ram energy. The upstream density of these ions closely follows the magnetic field, which forms a large loop whose both ends are located at the shock at two different locations in $z$, and the flux of suprathermal ions is nonuniformly distributed over the shock surface. The conclusion \cite{Giacalone2005b} draws is that the ions are specularly reflected and subsequently accelerated in these loops bouncing back and forth along the magnetic field from the shock. They thus do not need an upstream reflection; since they move along the loop magnetic field they are automatically returned to the shock between the bounces which accelerates them because if moves upstream in $x$ direction and because the particles during their many bounces experience the perpendicular upstream motional electric field that accelerates them in the direction perpendicular to the magnetic field and the flow. It is interesting to remark that this mechanism should in principle be independent of the presence of an upstream turbulence as the reflected ions anyway experience this acceleration when participating in perpendicular shock reformation. If there are many reflected ions, which is the case at high Mach number, their thermal pressure should locally cause the expansion of the magnetic field into the upstream direction and form magnetic loops along the shock surface. The upstream turbulence and the self-excited waves in the shock foot amplify this effect until large loops are formed in the interiors of which the particles are temporarily trapped and accelerated to high energies. The left part in Figure\,\ref{chap6-fig-hepsim} shows the particle differential energy flux as function of energy normalised to the upstream flow energy. Two spectra are shown both taken in the interval $335<x/\lambda_i<490$ but for different at different  domain sizes $z_{max}=\lambda_{corr}$ in $z$. These flux spectra show the formation of a high energy population on the particle fluxes up to energies of 100 upstream flow energy. These spectra are very steep, however, and it requires very large correlation lengths and extension of the shock in order to flatten them substantially. Nevertheless, the value of this simulation is that it demonstrates the possibility of a perpendicular shock to accelerate particle to high energies under the condition that it is high Mach number and magnetic loops form in its upstream part either because of prevalent turbulence (as assumed in this simulations) or because of any other reasons as, for instance the working of instabilities like the Weibel instability or particle trapping which may cause expansion of the upstream magnetic field until forming loops. The importance of this mechanism is that it is capable of accelerating shock-reflected ions which are citizens of the thermal population.

\begin{figure}[t!]
\centerline{\includegraphics[width=1.00\textwidth,clip=]{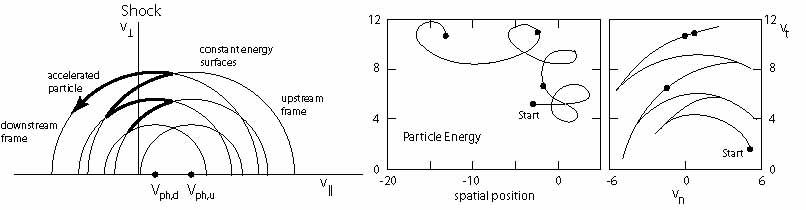}}
\caption[Sugiyama]
{\footnotesize Particle injection from a quasi-parallel shock by trapping acceleration across the shock \citep[after][]{Sugiyama2001}. {\it Left}:  Schematic of the mechanism. The upstream and downstream waves move at different phase velocities. The curves show circles of constant energy for a trapped particle. The ion moving back and forth across the shock along the different lines picks up the energy between the upstream and downstream wave frames exactly like in Fermi acceleration, until having sufficient energy to escape and enter the real Fermi process. {\it Right}: Particle orbits in this process. The energy increases while the particle is moving with the waves. The outer panel shows the bouncing in normal and tangential velocity coordinates.} \label{chap6-fig-terasawa}
\end{figure}

\subsubsection{Terasawa's mechanism of trapped particle acceleration}
\noindent In an attempt of finding a mechanism that explains the extraction of ions from the thermal particle population in the interaction with the shock inferred by \cite{Scholer1990}, \cite{Kucharek1991}, \cite{Scholer1992} and \cite{Scholer1999} from their hybrid simulations, \cite{Sugiyama1999} suggested a simple mechanism which could efficiently accelerate ions trapped in a large amplitude upstream wave when the wave is crossing the shock. The mechanism exploits the fact that the flow velocity of an upstream (monochromatic) wave decreases when the wave is crossing the shock. This is obvious when remembering that the velocity of upstream waves is essentially the Alfv\'en speed. During the time of passage of the wave across the narrow shock transition, the trapped particle in this case bounces back and forth across the shock transition between a fast and a slow propagating wave and has a chance to pick up the phase velocity differences. The particles moves along the lines of constant energy when jumping from one frame to the other. The process gradually increases the particle energy. For a high bounce frequency of the particle it experiences many reflections and can thus be speeded up to an energy which is high enough for injection into the Fermi mechanism. The action of the mechanism is schematically shown for a monochromatic wave in Figure\,\ref{chap6-fig-terasawa}. This mechanism has been elaborated in a paper by \cite{Sugiyama2001} of which the particle orbit and energy gain is shown on the right in the figure. The process is essentially diffusion in energy space which for sufficiently large wave amplitudes becomes chaotic.\index{acceleration!trapped particle}

A restriction of this process is the assumption of monochromatic waves. Simulations have shown that quasi-parallel shocks are built of large amplitude and irregulary shaped pulsations. These pulsation are fast before becoming the shock, i.e. before reaching the shock. It is also seen in the simulations that ions are not only retarded but also trapped between the pulsations and the old shock front. During the approach of the pulsation they may bounce many time between the old and new shock, a process that is very similar to Terasawa's wave acceleration and that might accelerate a fractin of the trapped ions to high energies.

\section{Accelerating Electrons}\index{acceleration!electron}
\noindent Unlike in astrophysics, shock acceleration of electrons is of secondary importance in the Heliosphere. The first and obvious reason -- with few exceptions -- is the accessibility of most of the shocks in the Heliosphere which allows for the performance of measurements {\it in situ} the shock environment. The second and less obvious reason is that in non-relativistic collisionless shocks the dynamics of the shock, shock formation and re-formation is mainly governed by ions which are the main carriers of the kinetic energy in the primary supercritical Mach number flow. This is also the reason for the great success of hybrid simulations in describing the shock dynamics. 

It has only recently been realised that electrons do also affect the shock dynamics at high Mach numbers through the excitation of instabilities in which ions are involved and  through small scale currents flowing on the microscopic scale $\sim\lambda_e$. Otherwise electrons serve the remote sensing and diagnostics of shocks from  remote distances. Such inaccessible shocks are the shocks in the solar atmosphere, from the chromosphere up into the high corona, long before some of the shocks that have been generated there manage to escape into interplanetary space and becoming visible as CME driven shocks, shocks in corotating interaction regions (CIR shocks) and, closest to the sun, the solar type II burst shocks. 

As long as the shocks remain to be hidden in the hot corona where spacecraft cannot enter, such shocks are realised only in the signature of the radio radiation they emit, as well as in X-rays and sometimes even Gamma-rays. This radiation is commonly used for diagnostical purposes that are based on theories of its excitation. Here the acceleration of electrons becomes of interest as well, even though some of the observed accelerated electrons and their signatures in the radiation are definitely not shock generated. The exempt applies to solar type III radio bursts and possibly also to some classes of prompt solar X-ray and Gamma-ray flares, which are believed to be caused by reconnection rather than shock acceleration, or by the direct action of localised strong electric potential drops along the local magnetic fields. In the following we, however, restrict ourselves to a concise description of what is currently known about the acceleration of electrons by non-relativistic collisionless shocks. 

\subsection{Introductory remarks}
\noindent Electron acceleration in non-relativistic shocks meets at least two severe problems. The first problem is that a super-critical shock must reflect ions in order to get rid of the excess energy, simply because ions at a given Mach number carry the bulk of the kinetic energy. In order to do this the shock potential is positive. A simple explanation for the generation of a positive potential in a high Mach number flow at the quasi-perpendicular shock transition can be based on the large discrepancy between the convective electron and ion gyroradii: at same perpendicular speed we have $r_{ce}/r_{ci}=m_e/m_i$, with the consequence that ions penetrate considerably deeper into the shock transition than electrons. A similar condition holds for the inertial lengths, which lead to a less restrictive ratio $\lambda_e/\lambda_i=\sqrt{m_e/m_i}$. The difference in penetration depth generates a positive space charge over a fraction of the distance $r_{ci}$ (or $\lambda_i$) and give rise to an upstream directed electric field component which accelerates electrons downstream and inhibits electron reflection. 

The second problem is also related to the small electron gyroradius (or inertial length) because it restricts the shock normal angle of electron reflection to near perpendicular. A third difficulty is that for electrons in a stream like the solar wind the kinetic energy of the streaming motion $\frac{1}{2}m_eV^2\ll T_e$ is usually much less than the electron thermal energy. In fact, for a streaming velocity of $V=1000\,{\rm km\,s}^{-1}$ the kinetic energy is just $\sim 5$\,eV, while the solar wind electron temperature is $T_e\simeq (50-100)$\,eV. Even though the shock is supercritical,  the dominant electron speed is the electron thermal velocity. Writing $r_{ce}^2/r_{ci}^2=(m_e/m_i)(T_e/K_i)$, with $K_i=m_iV_i^2/2$ the ion kinetic energy, we have $r_{ce}/r_{ci}=\sqrt{T_e/K_i}\,(\lambda_e/\lambda_i)$. In the solar wind $K_i\sim $\,1\,keV. Hence, due to the large electron temperature, $r_{ce}/r_{ci}\sim \sqrt{m_e/10\,m_i}$ is just a factor 3 smaller than the ratio of the inertial lengths. However, the high electron temperature means that the electrons are about isotropic with negligible flow speed, and flow-dependent electron reflection {\it per se} as considered in the next section does practically not depend on the flow.

\subsection{The Sonnerup-Wu mechanism}\index{process!mirror reflection}\index{Sonnerup, B. U. \"O.}\index{Wu, C. S.}
\noindent A simple reasoning based on a first attempt by \cite{Sonnerup1969} was put forward by \cite{Wu1984} in order to explain electron reflection from a nearly perpendicular shock. In a purely kinematic picture, the energy gain of a shock reflected particle depends on the de Hoffmann-Teller velocity
\begin{equation}
{\bf V}_{\rm dHT}=\frac{{\bf n\times V}_1\times{\bf B}_1}{B_{1n}}, \qquad B_{1n}\equiv {\bf n\cdot B}_1
\end{equation}
where ${\bf n}$ is the shock normal vector. The shock velocity along the upstream magnetic field is $V_{s\|}=V_{\rm dHT}\sin\thetabn+V_{1\|}$. We recall that the above definition of ${\bf V}_{\rm dHT}$ makes sure that in the de Hoffmann-Teller frame the bulk upstream flow has no component perpendicular to the upstream magnetic field. The complete upstream flow is along ${\bf B}_1$ in this frame, i.e. the de Hoffmann-Teller frame translates in the perpendicular direction with the velocity of the magnetic field lines projected to the shock surface. However, at a nearly perpendicular supercritical shock the parallel flow speed $V_{1\|}$ is small. In the expression for $V_{s\|} $ it is then negligible.\index{frame!de Hoffmann-Teller} 
\begin{figure}[t!]
\centerline{\includegraphics[width=1.00\textwidth,clip=]{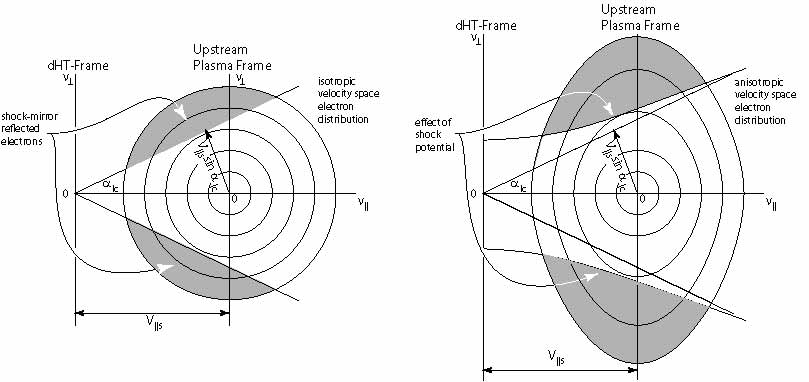}}
\caption[Wu]
{\footnotesize Electron phase space distribution in velocity space seen from two frames, the upstream plasma frame and the de Hoffmann-Teller frame, on the left the isotropic distribution \citep[after][]{Wu1984}, on the right the anisotropic distribution with $T_\perp>T_\|$. The dHT-frame is obtained translating the parallel flow velocity in the direction along the upstream magnetic field by the parallel shock velocity $V_{\|s}$. The loss-cone angle seen from the dHT-frame cuts out the particles of low speed which pass the shock. The higher energy particles (shaded) are `trapped' in the upstream flow magnetic field and can become reflected. On the right the enlarging effect of the attractive ambipolar electric shock potential on the loss cone is shown schematically. For being mirror reflected, electrons require quite a large temperature or perpendicular anisotropy.} \label{chap6-fig-wu1}
\end{figure}

Considering the particle distribution in the de Hoffmann-Teller frame is achieved by subtracting $V_{s\|}$ from the parallel upstream particle speed $v_\|$. Assuming that a particle can be mirror reflected at the shock under conservation of the particle magnetic moment implies that the particle velocity measured in the de Hoffmann-Teller frame has a pitch angle against the magnetic field that is larger than the loss-cone angle $\alpha>\alpha_{lc}$ defined through $\sin^2\alpha_{lc}=B_1/B_{os}$, where $B_{os}$ is the overshoot (or maximum) magnetic field in the quasi-perpendicular shock transition. This is shown schematically in Figure\,\ref{chap6-fig-wu1}. This loss cone angle is completely determined through the compression ratio of the shock. For a compression ratio of 3 the loss cone angle is $\alpha_{lc}=35^\circ$ covering a large part of the distribution function for large de Hoffmann-Teller speeds corresponding to large $V_{s\|}$. The narrowest loss cone would be obtained for the largest (fluid) compression ratio 4, yielding just $\alpha_{lc}=30^\circ$. These numbers are based on the Rankine-Hugoniot MHD relations and do not account for the overshoot magnetic field, which narrows the loss cone angle a little more.

Since most of the particles are not reflected but pass the shock, these particles are lost from the distribution, which is equivalent to the fact that particles that can possibly be reflected must already have sufficiently high velocity, a conclusion having been drawn already by \cite{Fermi1949}. One is thus forced to assume that the electron distribution consists of a thermal core distribution and a superposed halo distribution $F_e(v)=F_{core}+F_{halo}$. The number density of reflected particles follows from
\begin{equation}
N_{\rm refl}=2\pi\int_0^\infty {\rm d}v_\| \int_{v_\|\tan\,\alpha_{lc}}^\infty v_\perp {\rm d}v_\perp F_e(v_\|,v_\perp)
\end{equation}
where the distribution has to be transformed into the de Hoffmann-Teller frame. Clearly, according to the above assumptions, only the halo distribution contributes to this integral. In this formulation the electrostatic ion-reflecting field in the shock ramp is not included. This can however be corrected as has been shown by \cite{Goodrich1984}. The effect of the electrostatic shock potential is to accelerate the impacting electrons in the direction downstream of the shock. This widens the electron loss cone and reduces the number of reflected electrons further. 

As long as sufficiently high energy electrons are present in the upstream plasma some of them will become reflected due to the action of the `shock mirror force' and will return into the upstream flow where they experience the upstream electric induction field and become accelerated in the direction opposite to the reflected ions thereby increasing the current in the foot of the quasi-perpendicular shock. (Note that this amplification of the shock foot current further increases the vulnerability of the shock foot to the excitation of various current driven instabilities like the Buneman two-stream and modified two-stream instabilities.) This perpendicular acceleration further increases the electron energy and thus increases the chance of the electrons to become a second time reflected. In addition it introduces an anisotropy (see the right part of the figure) into the accelerated electron distribution with higher perpendicular than parallel energy, elongating the electron distribution into the perpendicular direction. Again this is in favour of the reflectivity but also may serve as a source for the whistler instability with the consequence of enhanced pitch angle scattering and progressing isotropisation.  Hence, for a small number of electrons multiple reflection is programmed into this mechanism of reflection. 

\begin{figure}[t!]
\centerline{\includegraphics[width=0.60\textwidth,clip=]{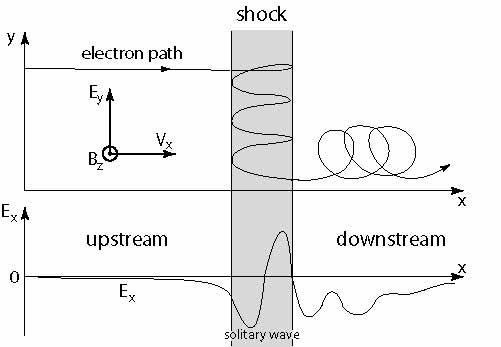}}
\caption[Shimada-Hoshino ApJL]
{\footnotesize Sketch of Hoshino's electron shock surfing mechanism \citep[after][]{Hoshino2001}. The electron after arriving at the shock is trapped for a while inside the shock potential which in this case is assumed to be a solitary waves. After having become accelerated the electron leaves the shock downstream. }\label{chap6-fig-sketchSS}
\end{figure}
\subsection{Hoshino's electron shock surfing mechanism}\index{process!shock surfing, electrons}
\noindent This magnetic mirror reflection mechanism completely neglects any electric potential drop in the shock transition. There are several contributions to such a potential drop. The first is due to the inertia difference between ions and electrons, which induces the general ambipolar large-scale shock electric field which we have discussed earlier in relation to ion shock surfing.  Electrons, however, are attracted by this potential and will not be reflected. (This has been noted above in passing when speaking about the shock electric field effect which diminishes the magnetic mirror effect.) However, the intensification of the current in the shock front by the mirror effect can cause several instabilities to grow and can substantially modify the structure of the electric field such that electrons can under certain conditions become reflected, thus intensifying the mirror reflection process, an idea that has been made use of in the \cite{Hoshino2001} electron shock acceleration model and the related simulations by \cite{Hoshino2002} and \cite{Amano2007}.  

\subsubsection{Electron shock surfing mechanism}\index{shocks!electron surfing}\index{acceleration!electron surfing}\index{process!shock surfing, electrons}
\noindent The fact that the passing electrons -- i.e. the non-reflected part of the upstream plasma electron population that crosses the shock for afterwards constituting the downstream plasma electrons -- is heated and partially  thermalized by the action of micro-instabilities in the shock transition layer when crossing the shock led  \cite{Hoshino2001} to propose that these micro-instabilities might be capable not only of thermalising the plasma but also of accelerating particles (ions and electrons) to large energies (as sketched on the right in Figure\,\ref{chap6-fig-cartoon}, and detailed in Figure\,\ref{chap6-fig-sketchSS} electrons). Since is is the electrons which are most vulnerable to micro-instabilities this acceleration would be of particular interest in electron acceleration. In Hoshino's words, these ``non-thermal accelerations that occur in the shock transition layer/shock front layer are still controversial, but at least they are believed to serve as `seed' particles which are subject to further acceleration to much higher energies by the diffusive shock acceleration" mechanism. In contrast to diffusive acceleration which dwells on multiple shock crossings and reflections ``the acceleration in the shock front region is provided when the particle traverses a thin shock layer, and the acceleration time may be shorter than that of the diffusive shock acceleration" which therefore ``may be called (a) `fast' process". Justification for this assumption \cite{Hoshino2001} seeks from the recent observation of large amplitude highly localised electric fields in  shock transitions in space \citep{Matsumoto1997,Bale1998}, measurements that have been confirmed by later observations \citep{Bale2002,Behlke2004,Hull2006,Oka2006,Bale2007} and have been referred to in previous chapters. According to these observation the shock transition is a region where large amplitude solitary structures or BGK modes are generated. 

The most recent measurements by the Polar spacecraft \citep{Bale2007}\index{spacecraft!Polar} during a bow shock transition suggest parallel electric fields  $E_\|\lesssim 100\,{\rm mV/m}$, and  perpendicular electric fields $E_\perp\lesssim 600\,{\rm mV/m}$ on parallel scales comparable to the electron inertial length $\lambda_e$. Such fields correspond to localised electrostatic potentials of several $10$\,V  along the magnetic field and $\lesssim 1$\,keV perpendicular to the ambient shock magnetic field. Production of these structures requires the inclusion of electron dynamics and refers to nonlinear kinetic plasma theory. It is thus quite natural to assume that these structures are accompanied by electron acceleration and heating, in particular as on these short scales the electron magnetic moment is not conserved anymore and the electrons become effectively non-magnetic and vulnerable to prompt acceleration in the parallel and also in the perpendicular electric fields.

Considering the presence of the small-scale electric field structures, which are solitary structures of the family of BGK modes of spatial scales $\lambda_e\sim (10-100)\lambda_D$ several Debye lengths long, has two additional interesting aspects. 

The first aspect is that these solitary structures expel electrons from their interiors, i.e. they represent localised negative potentials which act repulsing on the electrons. Hence, while the charge separation in the shock front that has been mentioned previously to cause a problem in electron reflection, accelerates electrons downstream, the localised solitary structures compensate and overcompensate for this effect. Electrons can in this way become effectively reflected from the shock in spite of the shock ramp bearing a larger scale electron-attracting potential. 

The second aspect is that electrons have a chance of being reflected only when impacting on one of these solitary structures and feeling the solitary wave field. It is just the fraction of upstream electrons that collide with the solitary structures that is reflected. This explains why not all particles of smaller than maximum reflection energy will be reflected from the shock ramp. The particle should in addition collide with the solitary structure. For efficient acceleration it should also become trapped. This is possible in principle in two was: firstly, by entering the solitary structure in which it can become trapped because these BGK modes are positively charged and lack electrons inside. The electron for this to happen must overcome the negative wall around the BGK mode (which is in principle a quadrupolar structure built of the bipolar interior and the reflecting negative electron wall around; often only the bipolar structure is seen when the wall electrons are convected away). Secondly, by jumping from one solitary BGK structure to the next being from some limited time in spacetime coherence with the BGk modes.   

As Figure\,\ref{chap6-fig-sketchSS} suggests, both types of these reflected/trapped electrons experience the upstream motional electric field and become accelerated along the shock front in the direction opposing the direction of acceleration of reflected ions. Again, this increases the reflected particle current along the shock front. This current is thus striated consisting of a distributed shock foot current which contains much stronger current filaments in those narrow regions where the reflected electrons flow along the shock. The energy gained by the electron can be estimated from the electron equation of motion $\dot p_x=-eE_x-ev_yB_z$, where $p_x=m_e\gamma v_x$ is the electron momentum, and $E_x$ is the amplitude of the electrostatic BGK mode i the shock transition region. The trapped electron is accelerated  as long as the electric force $eE_x> ev_yB_z$. During the nonadiabatic phase the electron experiences when feeling the $E_y$ motional field, the velocity $v_y$ increases quickly until this inequality inverts and the electron escapes from the BGK mode soliton. The energy it can reach in this case follows from the condition that the two forces have equal magnitude, or $v_y=E_x/B_z$. Estimating this quantity requires knowledge of the solitary amplitude. Since in the Buneman two-stream instability case the BGK modes take their energy from the electron-ion current
\begin{equation}
\epsilon_0E_x^2\sim \zeta m_eNv_d^2, \qquad v_d\sim 2V_1
\end{equation}
where $v_d$ is the current drift velocity which is the difference between the reflected ions and the inflowing electrons which is assumed to be responsible for the Buneman instability. This expression follows from equating the electrostatic energy in the BGk mode to the current drift energy (note that the current is carried by the electrons). The factor $\zeta$ is the conversion efficiency which from Buneman instability theory is taken as $\zeta\sim\frac{1}{4}\approx(m_e/m_i)^\frac{1}{3}$. On defining the upstream motional field $E_u=V_1B_1$ the above expression yields for the BGK structure amplitude
\begin{equation}
\frac{E_x}{E_u}=\frac{2c}{V_{1A}}\left(\frac{\alpha m_e}{m_i}\right)^{\!\!\frac{1}{2}}\simeq \frac{2c}{V_{1A}}\left(\frac{m_e}{m_i}\right)^{\!\!\frac{2}{3}}
\end{equation}
In the solar wind the Alfv\'en velocity is between $50<V_{1A}<150\,{\rm km/s}$  yielding $15<E_x/E_u<45$, which corresponds to solitary wave amplitudes of $E_x < 500\,{\rm mV/m}$ well in the range of observations in near Earth space. 

The maximum velocity that the electron can attain is then obtained from the equilibrium between the electric and Lorentz forces yielding $v_y^{\rm max}\lesssim 2{\cal M}_Ac(m_e/m_i)^\frac{2}{3}$ which corresponds to a maximum energy
\begin{equation}
\frac{{\cal E}_e^{\rm max}}{m_ec^2}\lesssim 2 {\cal M}_A^2\left(\frac{m_e}{m_i}\right)^{\!\!\frac{4}{3}} \sim 10^{-4}{\cal M}_A^2
\end{equation}
This estimate shows that for medium large Mach numbers of the order of, say, ${\cal M}_A\sim 10$ electrons can gain a maximum energy of a few per cent of their rest energy, which is in the $\sim10\,\, {\rm keV}$ range. This should in principle be sufficient for entering the Fermi cycle. On the other hand, for very large Mach numbers the possibility arises that the electron remains trapped for very long time in the BGK solitary structure, in which case the inequality is inverted, and the electron may gain energy quite far above its rest energy. This will happen for Mach numbers ${\cal M}_A> 100$, cases that are realised in astrophysical systems.\index{energy!gain,electrons maximum}

\begin{figure}[t!]
\centerline{\includegraphics[width=0.60\textwidth,clip=]{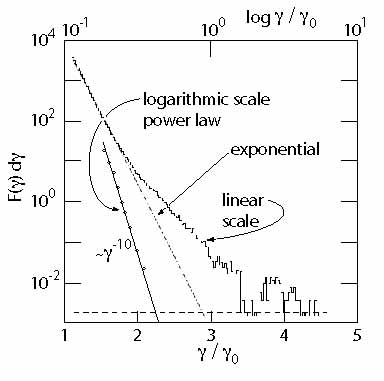}}
\caption[Shimada-Hoshino ApJL]
{\footnotesize A high Mach number (${\cal M}_A=32$) electron shock surfing PIC simulation for quasi-perpendicular shocks, showing the downstream electron distribution \citep[after][]{Hoshino2002}. The original data are given on linear energy scale (bottom abscissa) and rescaled into logarithmic energy scale (top abscissa). The dash-dotted line in linear scale indicates an exponential. The rescaled logarithmic curve is linear in log-log and therefore represents a power law which is quite steep, $\propto \gamma^{-10}$. The electron spectrum is exponential at low and power law at high energies. }\label{chap6-fig-shim}
\end{figure}

\subsubsection{Shock surfing simulations}
\noindent Several increasingly sophisticated numerical simulations have been performed in order to check the proposal of the electron shock surfing mechanism. The question arises which instability is responsible for the generation of the solitary structures which could reflect the electrons. 
There are a number of instability candidates which we have discussed in the chapter on quasi-perpendicular shocks. These are the Buneman two-stream instability, the modified two-stream instability, and the ion sound instability. 

The Buneman instability requires electron thermal speeds smaller than the electron-ion current drift speed $v_d=v_{di}-v_{de}\approx -v_{de}>v_{e{\rm th}}$. Taking the reflected ions and incoming electrons it can marginally grow with growth rate $\sim 0.03\omega_{pe}$. It immediately heats the electrons until it interrupts its growth by letting the drift velocity drop below thermal speed. It, however, quickly forms the required BGK modes which reflect and also trap low energy electrons. On the other hand the electrons  heated by the Buneman instability may lie outside the de Hoffmann-Teller loss cone in which case they can become mirror reflected. These electrons are further accelerated in the foot thereby increasing the foot current until it drives another instability. \index{waves!BGK modes}\index{waves!solitary structures}

However, \cite{Matsukiyo2003,Matsukiyo2006} have shown by particularly tailored full particle PIC simulations that the electron current driven modified two stream instability is a stronger instability, growing faster than the ion cyclotron frequency, and in addition has a lower threshold than the Buneman two-stream and ion acoustic instabilities. It heats the electrons parallel to the magnetic field and produces BGK phase space holes. The situation is quite complex switching between the different reflected and inflowing species, their densities and temperatures and the angles to the magnetic and electric fields. Only simulations can help understanding the acceleration of electrons.

Figure\,\ref{chap6-fig-shim} shows the final electron distribution function (differential energy flux) as the result of a high Mach number (${\cal M}_A=32$) full particle electron shock surfing PIC simulation for a quasi-perpendicular shock. At this high Mach number restriction required that the mass ratio was set to $m_i/m_e=20$, fairly unrealistic. Since the electrons have lower mass they must be treated relativistically in such simulations. Hence $\gamma={\cal E}/m_ec^2$ is the energy. The distribution is given on two scales, showing the evolution of a long nonthermal tail. Note that the dotted line in the linear scale is the corresponding Maxwellian. At energies $\gamma/\gamma_0<2$ the distribution is exponential ($\gamma_0$ is the relativistic energy of the incident flow electrons). At higher $\gamma$ it deviates becoming non-exponential. The log-log representation shows that the high energy tail is clearly power law $\propto\gamma^{-\alpha}$ with power $\alpha\sim 10$. This is a very steep power law, however. Nevertheless it shows that electron can be accelerated. \cite{Hoshino2002} trace this acceleration back to the action of the Buneman instability in their case. The problem is that at this low mass ratio indeed the Buneman instability is dominant, while at realistic mass ratios the Buneman instability becomes unimportant. This has been shown by \cite{Matsukiyo2006}. It might be the reason for the obtained results.\index{threshold!Buneman}
\begin{figure}[t!]
\centerline{\includegraphics[width=0.95\textwidth,clip=]{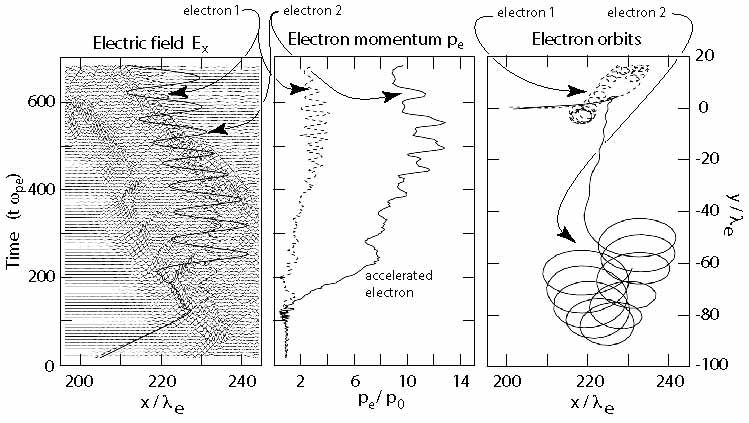}}
\caption[Hoshino-Shimada ApJ]
{\footnotesize High Mach number (${\cal M}_A=32$) electron shock surfing PIC simulation for quasi-perpendicular shocks, showing {\it Left}: the time-stacked evolution of the electric wave field $E_x$ in the shock transition with two electron orbits overlaid. The dashed orbit belongs to an electron not in resonance with the wave, the solid line is the resonant electron. In contrast to electron 1, electron 2 performs large excursions around the shock position before leaving the shock. Electron 1 also moves with the shock but crosses it without gaining energy. {\it Center}: Evolution of the total momentum of the two electrons in time. Electron 1 gains very little momentum/energy after entering the shock, while electron 2 Initially gains very much momentum and then enters a nonlinear state where the gain is slower. {\it Right}: The two electron orbits in the $(x,y)$-plane. Electron 1 moves only a small distance in both $x$ and $y$, while electron 2 performs a long initial jump in $y$ at constant $x$ after which it becomes trapped in the wave and bounces back and force with its enlarged gyroradius \citep[after][]{Hoshino2002}. }\label{chap6-fig-shim1}
\end{figure}

\cite{Hoshino2002} in their one-dimensional PIC simulation have shown that in interaction with one single BGK mode the electron cannot gain more energy than $\gamma/\gamma_0=1.26$. Hence, the acceleration to higher energies requires a mechanism of shock surfing, i.e. further interaction in the motional electric field and possibly secondary reflections. The BGK modes serve mainly the reflection of electrons including some pre-acceleration. The path of an electron along the shock is in these simulations shown in Figure\,\ref{chap6-fig-shim1}. This figure, on the left shows the time evolution of the shock normal electric field component $E_x$. This field exhibits a quasi-periodic structure in the foot of the shock showing that the field is concentrated in narrow spatial regions which move in a characteristic way first in upstream direction, turn around and propagate n downstream direction until decaying away. Two selected particle orbits have been superimposed on the field, one of the particles not being affected by the presence of the electric field (labelled electron 1), and another particle that interacts with the wave electric field (labelled electron 2). 

Both electrons enter the shock with the upstream convective flow velocity along the straight line at the bottom of the figure. Electron 1 does not feel the wave electric field because of some obscure reason when encountering the shock. It passes over the crest of $E_x$. In the back of the electric field it starts performing an oscillatory motion as it is now stopped becoming a slowly moving member of the downstream flow.  The central panel shows that its energy gain is small due to adiabatic heating only, while the right panel shows that it has been moderately heated in perpendicular direction performing its gyrational motion.

Unlike electron 1, electron 2 when encountering the repelling electric field around about time $t\omega_{pe}\sim 100$ is stopped in its downstream directed motion and has become slightly reverted into $-x$ direction. For roughly 100 plasma periods it stays trapped in the electric field and follows its path in upstream direction. The central panel shows that during this time is becomes violently accelerated about eightfold times. Having gained that much energy it manages to escape the field and pass to downstream performing now large oscillation in $x$. The two-dimensional representation of its orbit shows that similar to electron 1 it has escaped from the electric field but has shifted a long distance along the shock in $y$ direction. The oscillations in $x$ seen in the left panel unmask as being projections of its gyro-orbit showing that electron 2 has been accelerated nonadiabatically perpendicular to the magnetic field during its contact with the localised electric field. The lack in acceleration of electron 1 can thus be interpreted that electron 2 has encountered the shock at a position $y$ without wave electric field. The electric field must hence have been highly localised. In fact,  from the right panel one realises that electron 1 encountered the shock at $y\sim 10\lambda_e$ only. It is also interesting to see from the right panel that both electrons have not moved far downstream during the time interval of this simulation shown. They are hanging around at a fixed location in $x$ while the shock is moving in the direction upstream. It is only in the shock frame that they have become displaced downstream of the shock.  

\begin{figure}[t!]
\centerline{\includegraphics[width=1.0\textwidth,clip=]{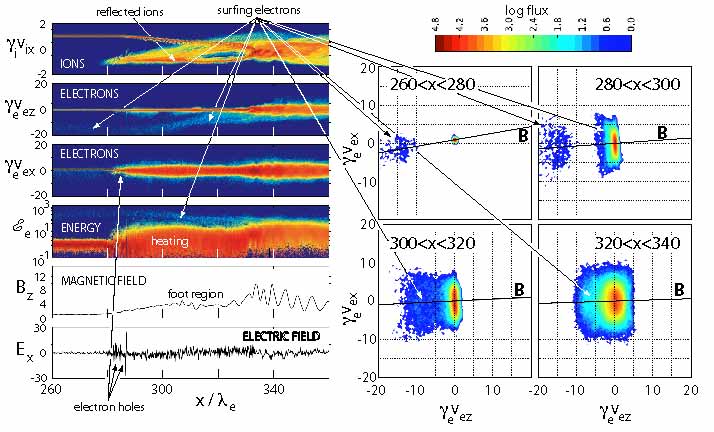}}
\caption[Amano1]
{\footnotesize One-dimensional electromagnetic full particle PIC simulation of high Mach number (${\cal M}_A=15$) electron shock surfing  for quasi-perpendicular shocks ($\thetabn=80^\circ, \beta_i=\beta_e=0.08$) at time $t\omega_{pe}=12000$, corresponding to $t\omega_{ci}=5.5$ \citep[after][]{Amano2007}. The ion-to-electron mass ratio is $m_i/m_e=100$, and the ratio of electron plasma to cyclotron frequency is $\omega_{pe}/\omega_{ce}=20$ implying a very low plasma density. The initial magnetic field ${\bf B}_0=B_{0z}\sin\thetabn\,{\bf e}_z$ is in $z$ direction. {\it Left}: The figure shows from top to bottom the ion phase space $(v_{ix},x)$, two components of the electron phase space $(v_{ez},v_{ex},x)$, electron kinetic energy ${\cal E}_e=(\gamma_e-1)m_ec^2$, with $\gamma_e=(1-v_e^2/c^2)^{-\frac{1}{2}}$ and $v_e$ the speed of each electron in the simulation frame, magnetic field component $B_z$ and electric field component $E_x$. Velocities are normalised to the injection speed $V_1$, magnetic field to upstream field, electric field to $E_{0y}=V_1B_{0z}$, kinetic energy to $(\gamma_1-1)m_ec^2$, with $\gamma_1=(1-V_1^2/c^2)^{-\frac{1}{2}}$.  The spatial scales are normalised to the electron inertial length $\lambda_e=c/\omega_{pe}$. {\it Right}: Electron phase space distributions taken in different space intervals in the left part of the figure successively approaching the shock transition.  The direction of the local average magnetic field in the respective spatial range is shown as the straight line labelled ${\bf B}$. The colour coding of the particle counts as function of the velocity components is logarithmic as given in the bar on top of the figure. The relevant information about reflected ions, surfing and accelerated electrons has been indicated by light arrows. Note that for resolution of the weakly relativistic energetic electrons, the code has to be relativistic. Hence the velocities given here in both phase space representations are actually the components of the 4-velocities ${\bf v}_e\to \gamma_e{\bf v}_e$.}\label{chap6-fig-amano1}
\end{figure}

Considerable progress has been achieved recently in understanding the electron shock surfing process through high resolution one-dimensional fully relativistic full particle PIC simulations by \cite{Amano2007} of a quasi-perpendicular ($\thetabn=80^\circ$) high Mach number shock (${\cal M}_A= 15$, Alfv\'en velocity $V_A=0.05\,c$) and for an ion-to-electron mass ratio of $m_i/m_e=100$, upstream frequency ratio $\omega_{pe}/\omega_{ce}=20$ fixing the plasma density for a given upstream magnetic field, and plasma beta $\beta_i=\beta_e=0.08$. Note that for resolution of accelerated electrons a simulation should have to be relativistic. The mass ratio is still unsatisfactory when referring to the investigation of \cite{Matsukiyo2006} that the Buneman instability ceases to be important at realistic high mass ratios $m_i/m_e=1836$ in which case it is replaced by the modified two-stream instability. In addition, one-dimensionality of the simulation misses obliquely propagating waves and electrostatic structures and can thus be only approximate. Nevertheless, the new simulation confirms shock surfing as a viable mechanism of electron acceleration and reveals a number of additional properties of shock surfing.    

\subsubsection{Detailed electron dynamics}
\noindent Figure\,\ref{chap6-fig-amano1} gives an overview of the simulations at a late time $t\omega_{pe}=12000$ (corresponding to $t\omega_{ci}=5.5$). From top to bottom the panels show the incident ion phase space $(\gamma_iv_{ix})$, two cuts through the electron phase space $(\gamma_ev_{ez},x)$ and $(\gamma_ev_x,x)$, electron kinetic energy ${\cal E}_e=(\gamma_e -1)m_ec^2$, magnetic field $B_z$, and the shock normal electric field component $E_x$. All data have been normalised accordingly (see the caption). The uppermost panel shows the incoming, reflected and shock heated/thermalised ions. \index{particles!surfing electron dynamics}

The second panel from top shows the gyrating incoming electrons as well as  at negative velocities $v_{ez}<0$ the surfing electrons. The third panel shows the heating of the electron inflow and surfing electrons. The fourth panel is the electron energy showing the enormous electron heating in connection with the flow entering the shock foot region and interacting with the reflected ion component. It is also worth noting that in the three electron panels fluxes of highest energy electrons flowing along the shock and along the magnetic field in $-z$ direction can be identified at a distance far upstream of the shock foot region. The effect of the reflected ions is seen in the formation of the foot in the magnetic field. 

However, the most interesting panel is the bottom panel which shows the shock normal electric field component which in the shock ramp evolves from the ambipolar charge separation while in the shock foot is entirely due to the action of micro-instabilities. Large amplitude signatures of solitary structure BGK modes evolve close to the edge of the shock foot. Here the relative drift between reflected ions and inflowing electrons is largest. They are followed deeper in the foot by a multitude of smaller amplitude but still non-linear signatures of electrostatic turbulence. In the present simulation these structures evolve because of the growth of the Buneman two-stream instability, and we note that at a higher ion-to-electron mass ratio this picture would change somewhat because the modified two-stream instability would become more important than the Buneman instability. The other instability concurring with the Buneman mode is the ion-acoustic instability. Its growth rate is, however, of the order of $>10\omega_{ci}^{-1}$, too long for becoming important in the distant foot but presumably responsible for some of the heating in the shock transition region.

\begin{figure}[t!]
\centerline{\includegraphics[width=0.6\textwidth,clip=]{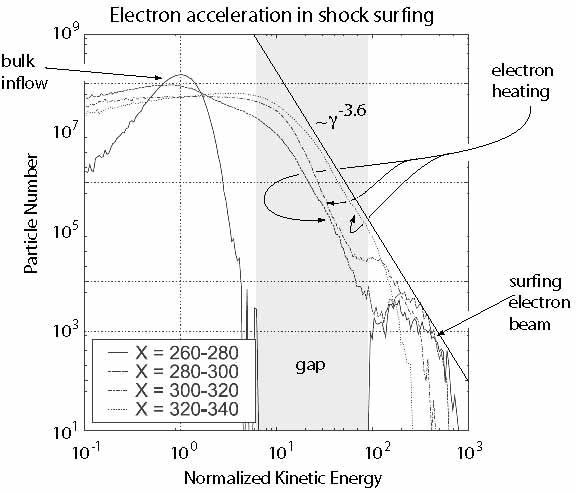}}
\caption[Amano2]
{\footnotesize Spatial evolution of electron energy spectra for the four spce intervals on the right in Figure\,\ref{chap6-fig-amano1} in electron acceleration by shock surfing \citep[after][]{Amano2007} . Progressing from the upstream foot boundary into the shock the electron spectrum widens with increasing electron temperature and in the centre of the shock evolves from an exponential into a distribution that over a certain energy interval exhibits a clear power law shape. Note the gap between the surfing beam and the bulk inflow distribution which is gradually filled by heating.}\label{chap6-fig-amano2}
\end{figure}

On the right in Figure\,\ref{chap6-fig-amano1} the electron phase space distribution is plotted in four spatial sectors from the edge of the shock foot right up to the shock ramp. At and outside the foot edge the distribution consists of the electron inflow and a dilute but very energetic upstream electron beam component which escapes at high negative speed along the magnetic field. Closer to the shock in the outer part of the shock foot the inflowing electrons have become heated perpendicular to the magnetic field and show a component of electrons that separates from the bulk electrons wanting to flow along the magnetic field in $z$ direction. In addition the surfing electron beam is seen at high but lower speed along ${\bf B}_1$ than in the former region at the foot edge. In the inner part of the foot the bulk electrons have become heated even more while a large group of electrons begins to completely separate from the bulk and to flow away along the field. These are the surfing electrons just during their first phase of acceleration. Finally, just at the ramp the electrons are substantially heated, but the distribution consists already of two parts, the bulk and another hot component at negative $v_{ez}$, the source population of the surfing electrons of which not all will participate in the surfing.

The electron energy distribution functions in these four regions are shown in Figure\,\ref{chap6-fig-amano2}.  The interesting property of these distributions is that the field aligned surfing beam seen most pronounced at the foot edge does survive through all the four upstream regions. It is forming a bump on the distribution function, while the bulk distribution is heated ever more with approaching the shock. The height and width of this bump varies slightly, but the beam electrons are still identifiable until close to the shock, and neither their energy nor their intensity does not vary strongly. This suggests a very fast acceleration mechanism. Moreover, the bumps occur only on the upstream distribution and are thus identified as shock reflected electrons. Since electrons of such energies have not been present initially in the original inflowing electron distribution, these bumps and the corresponding electrons cannot be the result of mirror reflection by the Sonnerup-Wu mechanism until the distribution has become heated as there have not been any particles outside the loss cone. For the Sonnerup-Wu mirror reflection mechanism the heating has to come first, and then the electrons can be reflected and accelerated in the upstream motional electric field. This acceleration is in the direction perpendicular to the magnetic field, however, while the electrons leave along the magnetic field. Note, however, that close to the shock transition the surfing electrons are heated almost isotropically (see Figure\,\ref{chap6-fig-amano1} on the right).

Heating of the bulk distribution is clearly seen in Figure\,\ref{chap6-fig-amano2}. The distribution widens, and between the beam electrons and the bulk maximum evolves into a region that exhibits an approximate power law shape with variable power law. The final shape in this simulation is found to have power $F_e(\gamma_e)\propto \gamma^{-3.6}$. However, it is clearly the surfing beam electrons that possess high energy of several $10^{\,2}(\gamma_{\,1}-1)m_ec^2$. For instance, in the solar wind with its bulk energy of ${\cal E}_{e1}\sim(\gamma_{\,1}-1)m_ec^2\sim 100$ eV, this corresponds to an energy ${\cal E}_e\sim\,{\rm of\,\, several}\, 10\,{\rm  keV}$, presumably sufficient for entering into the Fermi cycle for electrons. This can be concluded from a comparison of scales. For the Fermi mechanism the shock must be an infinitely thin surface. Electrons should thus have gyroradii $r_{ce}\gtrsim \Delta_s$. Since the shock width is of the order of $\Delta_s\sim 10^3$ km, this implies electron energies of ${\cal E}_e\gtrsim 50$ keV in a $B_1\approx 5$ nT magnetic field. In addition sufficient scattering of these electrons off upstream and downstream turbulence is required for further acceleration.

\begin{figure}[t!]
\centerline{\includegraphics[width=0.95\textwidth,clip=]{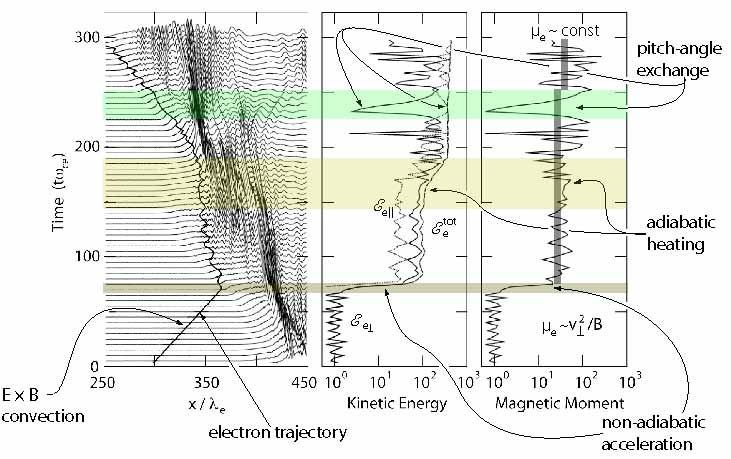}}
\caption[Amano2]
{\footnotesize Quasi-perpendicular shock electron acceleration by shock surfing \citep[after][]{Amano2007} {\it Left}: The time-stacked evolution of the magnetic field in the shock simulation plotted after $t\omega_{ce}=300\,(t\omega_{pe}=6000)$. Two cycles of quasi-perpendicular shock reformation (with period $\sim 2\omega_{ci}^{-1}$) are seen with the shock ramp jumping forward (to the left) against the upstream flow. The surfing electron trajectory (solid line) runs along the outer boundary of the shock foot. The electron is practically enslaved by the group of gyrating foot ions which are in excess of the flow and need to be charge neutralised.  {\it Center}: Evolution of the electron energy in perpendicular and parallel components and also the total electron energy. The strongest acceleration happens in the first encounter of the electron with the foot ion group. It experiences further acceleration each time the reformation cycle ends. Note the stronger perpendicular acceleration of the electron causing an electron anisotropy $T_{e\perp}>T_{e\|}$. {\it Right}: The same for the electron magnetic moment $\mu_e$ showing that the initial energy gain within a time interval of $\Delta t\omega_{ce}\sim 5$ is highly non-adiabatic with the magnetic moment changing drastically. Afterwards the average magnetic moment (grey bar) is constant. Note also the phase of adiabatic heating in the increasing magnetic field which is mainly in the perpendicular energy.}\label{chap6-fig-elsurf}
\end{figure}

Further insight into the surfing mechanism is again obtained by following a selected surfing electron path. This is done in Figure\,\ref{chap6-fig-elsurf}. In contrast to Figure\,\ref{chap6-fig-shim1} this time the left panel shows the particle trajectory superimposed on the stacked magnetic field instead of the wave electric field. This has the advantage to directly see the motion of the shock ramp and formation of shock foot during quasi-perpendicular shock reformation. 

The physics is however similar to what was concluded by \cite{Hoshino2002}. The particle moves in into the shock at the upstream convective flow velocity indicated by the straight line at the bottom. When encountering the shock foot edge containing the dense group of reflected gyrating ions the inflow motion of the electron is suddenly truncated, and the electron starts oscillating in $x$ for $\sim100\omega_{ce}^{-1}$ electron cyclotron periods around an almost stationary position. During this time the shock ramp approaches the electron at the speed at which the shock jumps ahead quasi-periodically during quasi-perpendicular shock reformation while the electron remains in the foot region not being able to cross the shock ramp. At the contrary, it surfs along the shock ramp and is taken over by the next cycle of reflected ions,  until the end of the simulation turning together with the shock reflected foot ions into the direction upstream away from the ramp in the formation of the next reformation cycle. 

We know of course that the electron is not being at rest at  position $x$. The electron moves in fact in the $-y$-direction along the shock being subject to acceleration by the upstream motional electric field. The excursions in its path in $x$ seen are the projections of the gyrations of the particle into the $(x,t)$-plane. Note that the injected electron had so low temperature that the gyrations remained hidden in the convective straight line electron path. The gyration becomes visible now because the electron has been violently accelerated at the encounter with the edge of the shock foot. 

This acceleration is seen from the central panel where the kinetic energy ${\cal E}_e=(\gamma_e-1)m_ec^2$ of the electron is plotted as function of simulation time. At the instant of the electron impact on the shock foot edge the electron energy suddenly increases by a factor $\sim10^2$. This acceleration is non-adiabatic which is indicated by the sudden change of the electron magnetic moment in the outer right panel at this time. The non-conservation of the first electron adiabatic invariant $\mu_e={\cal E}_{e\perp}/B$ signals that the electron is interacting on a scale shorter than the electron gyro-radius being effectively non-magnetic which can happen only when either the magnetic gradient is extraordinarily steep, which is not the case as seen from the left outer panel, or an electrostatic interaction takes place on a scale shorter than the electron gyro-radius. It can thus be attributed to large amplitude solitary BGK modes which are excited at the edge of the shock foot forming small scale electric potentials. This process is the same as in the simulations by \cite{Hoshino2002}.

From the central panel it is seen that the acceleration is predominantly in the perpendicular direction which is due to the trapping of the electron at the BGK mode where it comes to rest and experiences the motional electric field over the time of roughly $\sim 5\omega_{ce}^{-1}$ gyroperiods. After this time the magnetic moment does not change anymore until the next reformation cycle starts at time $t\omega_{ce}\sim 250$. The further increase in energy the electron experiences between $140<t\omega_{ce}<190$ is purely adiabatic caused by the electron moving up into the stronger ramp magnetic field which acts as a mirror and reflects the electron by the mirror force as proposed in the Sonnerup-Wu mechanism. Since the thermal upstream electrons cannot become reflected by no means by the Sonnerup-Wu mechanism, however, one realises that the pre-acceleration of the electron by the upstream convection electric field which is made possible by trapping the electron in the electrostatic BGK wave field, is a necessary condition for shock reflection of electrons and their further acceleration.  In between the two acceleration phases, the non-adiabatic and the adiabatic mirror reflection, there are some reversible variations in the magnetic moment without changing the particle energy that are accompanied by reversible exchanges between the parallel and perpendicular energies of the electron. They are induced by changes in the magnetic field direction (probably caused by phase locked whistler waves attached to the shock ramp) and related pitch angle variations which do not interest us in the context of electron acceleration.  

Applied to the solar wind the electron energy gained in the non-adiabatic phase is roughly $\sim 10$ keV, corresponding to a $\Delta\phi\sim 10$ kV electric potential drop. In a $B_1\sim 5$ nT field this yields a large acceleration rate of ${\rm d}{\cal E}_e/{\rm d}t\sim 1.5\,{\rm MeV/s}$. If the acceleration is due to the motional solar wind electric field of $E_{\rm sw}\sim 2.5$ mV/m, the acceleration length is of the order of $\ell_{acc}\sim 4\times 10^3$ km along the shock. Over this distance the electron remains to be in close contact with the BGK solitary wave. After leaving the BGK mode the accelerated electron has a large pitch angle against the magnetic field. It can now enter the Sonnerup-Wu mechanism for reflection at the shock since it will be found outside the loss cone which for a shock compression ratio of $B_2/B_1\sim 3$ is $\alpha_{lc}\approx 35^\circ$. 

\subsubsection{Digression on quasi-parallel shock surfing}\index{process!surfing, quasi-parallel shock}
\noindent From the above discussion of theory and simulation of electron surfing and acceleration at quasi-perpendicular shocks we  learn that it is the combined effect of the interaction between the reflected ion and incoming electron components, the generation of localised electrostatic solitary structures of the BGK mode family, electron heating and acceleration in the motional electric field, and the shock mirror effect that are responsible for the generation of high energy electrons at quasi-perpendicular shocks. Even though this combination is very complicated, it works only for quasi-perpendicular shocks, and it works the better the closer the shock is to being perpendicular as then the number of reflected ions at a given supercritical Mach number is highest and the mirror effect on the electrons is largest. This raises the question whether or not quasi-parallel shocks could as well reflect and/or accelerate electrons.

The question is difficult to answer. It has, however, been realised from hybrid and full particle simulations that quasi-parallel shocks reform due to the interaction of large amplitude magnetic pulsation. These pulsations can give rise to the excitation of high frequency electrostatic waves and broadband electric noise which can be understood only as the signature of many solitary like structures of the BGK type family. Observations by {\CL} have demonstrated the presence of such waves (see Chapter \ref{chap5-quasi-parallel shocks} on quasi-parallel shocks). Structures of this kind, as we know, are generated only when the electron dynamics is taken into account. Quasi-parallel shocks therefore seem to host a highly active electron dynamics. 

Moreover, during the  reformation of quasi-parallel shocks the direction of the shock normal angle fluctuates considerably, identifying the quasi-parallel shock in many cases as a wave dominated locally quasi-perpendicular shock on scales of the order of the tangential extension of the large amplitude pulsations or {\SL}. This evokes the conclusion that on those scales, locally, the same processes of electron reflection may take place as in the quasi-perpendicular case. Mirror reflection of electrons will then become possible. 

Ion reflection and foot formation happen along the front of a pulsation when the pulsation ({\SL}) approaches the shock, though on a smaller scale than in the extended quasi-perpendicular case. This might nevertheless be sufficient for electron surfing on the pulsation and subsequent acceleration. The whole effect could be extended over a much larger spatial region than in the quasi-perpendicular case where it is restricted solely to the shock foot region. In the quasi-parallel shock case it could be filling the entire region of presence of large amplitude pulsations in front of and at the ramp of the quasi-parallel foreshock. The related problems have not yet been properly considered and are open for future investigation.

\section{Conclusions}
\noindent Particle acceleration by shocks is just one aspect of collisionless shock physics even though for a large number of predominantly astrophysical applications it occupies the place of the most important property of a shock. Surprisingly, even after six decades of intensive research on this subject the original proposal made by Fermi  still survives in both of its versions, first and second order Fermi acceleration. However, considering the shock acceleration literature one cannot escape the impression that the whole field is stagnating for the last twenty years. The progress since the mid-eighties has been miniscule even though the number of papers has multiplied. That shocks accelerate particles, in the first place ions, is highly probable. It is supported by observations. Test particle simulations do support it as well. Reasons for obtaining power law particle spectra have been put forward, but the power laws obtained depend on the kind of theory and on the settings of the simulations. Theoretical reasons have been given for marginally flat power laws which in some cases seem to be in agreement with observation, but the field still looks erratic.  

While the basic acceleration mechanism for energetic ions by some kind of diffusive first-order Fermi acceleration will probably ultimately cover the problem, electron acceleration is still not  understood sufficiently well and, what concerns the acceleration of Cosmic Ray electrons,  will presumably be solved only when considering highly relativistic shocks. In addition, ion  acceleration still suffers from many open problems of which the injection problem is one of the most important. Analytical theory has so far been unable -- and will probably remain to be unable --  to provide a satisfactory or even just convincing answer to the question where the high energy ions originate that the diffusive Fermi acceleration mechanism needs to push them up to the observed hgh energies and the measured power law spectra. Bluntly said, the injection problem is too complex to be treated analytically. Lee's heroic self-consistent solution has a very limited range of applicability; test particle investigations ignore the effect of the shock itself and are valid only for very small numbers of accelerated particles which already possess high energies. In all of its versions even the most sophisticated which include different shock geometries, turbulence or radiation it covers just the generation of high energy tails on the distribution function and  do not contribute to the injection problem. 

On the other hand, self-consistent simulations of particle acceleration to high energies are still in their youth. The simulation boxes are too small to cover a sufficiently large spatial range and long enough simulation times; realistic mass ratios which have turned out to be of crucial importance must compete with requirements on resolution, box sizes, dimensionality and simulation time which all are restricted. Nevertheless, the injection problem will be solved only on the way of such self-consistent simulations. Leakage of ions from downstream to upstream is out of question. There is light over the horizon, however. For it seems that quasi-parallel shocks can accelerate ions until they overcome the injection threshold for entering into the Fermi acceleration cycle. These ions are produced by the shock self-consistently by accelerating them out of the bulk thermal distribution. The mechanism is still unclear being obscured by geometry, insufficient particle numbers, and low dimension. So far the acceleration of the required high energy ions has been seen only in one-dimensional hybrid simulations applying the tricks of energetic particle splitting while electron dynamics has been completely ignored. From quasi-perpendicular shock dynamics it has, however, become transparent that the structure, formation and space-time dynamics of the shock will be understood only when considering the realistic high ion to electron mass ratio. There is no reason  to believe that this would be different for quasi-parallel shocks.  

Therefore what is required are full particle PIC simulation at realistic mass ratios in three dimensions, large boxes and many particles. In the present chapter we have discussed several different aspects of the particle acceleration in one or the other approximation. Indeed deepest though quite preliminary insight into some of the problems has been obtained on the way of simulations  that were specially tailored to illuminate a particular problem like, for instance, the problem of perpendicular diffusion of energetic particles that have been accelerated under shock conditions. In this case it was found that the diffusion is weak while at the same time being time dependent and much faster than classical diffusion, a finding that in analytical theory has not yet been taken into account. Hybrid simulations applying energetic particle splitting, on the other hand, could demonstrate that there are good reasons to believe that quasi-parallel shocks are capable of pushing some particles over the Fermi threshold. Still, these simulations could not answer why this is so and what mechanism is responsible for this effect. Possibly it is the arrival of large amplitude upstream pulsations at the shock which cause a mirror effect on trapped ions when reforming the shock. Since electron dynamics is not considered in this case, nothing can be said about higher frequency waves, BGK modes, and solitary structures which have been identified in observations to exist in the shock transition region. 

The latter structures have turned out to be of crucial importance in quasi-perpendicular shock acceleration of electrons. Full particle simulations, still applying very low mass ratios and being performed in one dimension only, have convincingly demonstrated that it is the combined ion and electron dynamics that is responsible for acceleration of electrons by the shock surfing mechanism. Since it is known from quasi-perpendicular shock reformation that higher dimensions are of crucial importance and that full mass ratios modify the picture completely in many ways, not least by replacing the Buneman two-stream instability with the modified-two stream instability, even these simulations do probably not yet provide the realistic view. Nevertheless, in which however way the BGK modes in the shock foot are generated, by ion-electron or by electron-electron interaction, their presence is crucial for electron shock surfing and electron acceleration to high energies, electron beam formation at the shock and electron heating. 

At the quasi-parallel shock it is expected that similar mechanisms are at work simply because the upstream wave spectrum transforms the quasi-parallel shock locally on the small scale into a quasi-perpendicular shock, opening up the possibility of a similar process of electron acceleration also at quasi-parallel shocks. It is completely unknown what the effect would be on the ions, in particular as in the quasi-parallel foreshock the acceleration could proceed over a substantial part of the foreshock upstream of the shock. Self-consistent simulations are the only way to answer those questions.  Finally, the non-relativistic shocks treated here can contribute only to the understanding of the injection process. The system in the solar wind is not large enough and the upstream and downstream turbulence required to provide the scattering centres is not strong enough to shock-accelerate particles to extremely high cosmic ray energies.      


\end{document}